\documentclass{JHEP3}
 
\pdfoutput=1
\usepackage{graphicx}
\usepackage{eufrak}
\usepackage{slashed}
\usepackage{bm}
\usepackage{amsmath}
\usepackage{epsfig}
\usepackage{booktabs} 
\newcommand{\exclude}[1]{}
\newcommand{\ra}[1]{\renewcommand{\arraystretch}{#1}}

\title{Topological Currents in Neutron Stars: \\ \Large{Kicks, Precession, Toroidal Fields, and Magnetic Helicity }}
\author{James Charbonneau and Ariel Zhitnitsky\\ Department of Physics \& Astronomy, University of British Columbia \\
6224 Agricultural Road, Vancouver, B.C. V6T 1Z1, Canada\\ \\Email: james@phas.ubc.ca, arz@phas.ubc.ca}
\abstract{The effects of anomalies in high density QCD are striking.  We consider a direct application of one of these effects, namely topological currents, on the physics of neutron stars.  All the elements required for topological currents are present in neutron stars: degenerate matter, large magnetic fields, and parity violating processes.  These conditions lead to the creation of vector currents capable of carrying momentum and inducing magnetic fields.  We estimate the size of these currents for many representative states of dense matter in the neutron star and argue that they could be responsible for the large proper motion of neutron stars (kicks), the toroidal magnetic field and finite magnetic helicity needed for stability of the poloidal field, and the resolution of the conflict between type-II superconductivity and precession.  Though these observational effects appear unrelated, they likely originate from the same physics---they are all P-odd phenomena that stem from a topological current generated by parity violation.}

\begin{document}

\section{Introduction}
It is well known that anomalies have important implications for low-energy physics: the electromagnetic decay of neutral pions $\pi^0\rightarrow 2\gamma$ is a textbook example.  Anomalies reveal intricate  relationships between topological objects such as vortices, domain walls, Nambu-Goldstone bosons, and gauge fields and often result in very unusual physics.  A particularly relevant example is the superconducting cosmic string on which an electric current flows without dissipation and carries momentum  \cite{Witten:1984eb}.  The effects of anomalies are well established and are reviewed in \cite{Niemi:1984vz}.  

More recently the role of anomalies in QCD has been studied at finite baryon density \cite{Kharzeev:2004ey,Son:2004tq} and similar phenomena have been studied in condensed matter systems  \cite{RevModPhys.59.533,Alekseev:1998ds,Volovik2001195}.    Since these first steps many other applications of anomalies in  dense QCD have been considered: an analysis of the axion physics and microscopic derivation of anomalies \cite{Metlitski:2005pr,Metlitski:2005qz};  studying the vortex structure due to the anomalies currents in neutron stars (type-I versus type-II superconductivity) \cite{Charbonneau:2007db};  the charge separation effect at the relativistic heavy ion collider (RHIC) \cite{Kharzeev:2007tn,Kharzeev:2007jp};  magnetism of nuclear and quark matter \cite{Son:2007ny}; anomaly mediated neutrino-photon interactions at finite baryon density \cite{Harvey:2007rd}; the chiral magnetic effect at RHIC \cite{Fukushima:2008xe} and many others.  
  
For this paper we are interested in the result where a topological  vector current (along with an axial current) is induced in the background  of  an external magnetic field when the chemical potentials of the right and left-handed fermions are not equal, $\mu_r\ne \mu_l$.  As argued in \cite{Metlitski:2005pr} this phenomenon depends only on the presence of chiral symmetry and not on whether the chiral symmetry is spontaneously broken.    Applications of this induced vector current have been discussed in neutron star physics \cite{Charbonneau:2007db} and RHIC related physics \cite{Fukushima:2008xe}.  In  \cite{Charbonneau:2007db} it was shown that a non-dissipating vector current running along the magnetic flux tubes may change the behaviour of superconductivity from  type-II to type-I, even though the Landau-Ginzburg parameter $\kappa > 1/\sqrt{2}$ suggests type-II behaviour. It was also argued that the triangular lattice of flux tubes---the Abrikosov lattice---may be completely destroyed due to the helical instability in the presence of the induced vector current. This would  resolve the apparent contradiction with the precession of the neutron stars \cite{Link:2003hq,Link:2006nc}.  For another mechanisms that may   resolve this contradiction see \cite{Buckley:2003zf, Buckley:2004ca,Sedrakian:2004yq,Babaev:2009xv}.

Our goal is to present a quantitative analysis of the conditions when parity violation  ($\mu_r \neq \mu_l$) occurs in dense stars and a persistent, topological current is induced.  We claim that topological vector currents do exist in dense stars and we consider applications that may help explain many phenomena in dense stars, such as kicks and toroidal fields.  
  
 It is well known that the weak interactions (where parity is strongly violated) play a dominant role in neutron star physics. Producing the asymmetry $\mu_r \neq \mu_l$ for a given process is common in the bulk of neutron stars, but we are interested in coherent  parity violating effects  when the asymmetry  appears  in macroscopically large regions. Sections 3--7 are devoted to estimating the induced current  in different environments that may exist in neutron stars, while Section 8 is devoted to possible applications of the topological currents.  Readers interested in applications may go directly to Section 8, skipping the technical aspects of Sections 3--7. 
 
 If non-dissipating vector currents are induced they can transfer momentum either by escaping the star or radiating photons.  Our kick engine continues to work even when temperature drops well below $T \ll$ MeV.  This is because it is the chemical potential $(\mu_l - \mu_r)\neq 0$ that drives the kick, not the temperature.  This leads to long sustained kicks, rather than short natal kicks.  In Section 8 we use a rough calculation to demonstrate that our kick mechanism can explain the proper motion of hyperfast pulsars with $v > 800$ km/s, particularly B1508+55 moving at $v = 1083^{+103}_{-90}$ km/s \cite{Chatterjee:2005mj}.  A detailed calculation can be found in \cite{Charbonneau:2009hq}. See \cite{Lai:2003hm} for a review of kick velocities.   
 
 The kick naturally produces a magnetic field-momentum correlation $\langle  \vec{P}\cdot \vec{B} \rangle$.  Though it is believed that the spin and magnetic field are correlated somehow, the sustained nature of the kicks allows a rotation-kick correlation $\langle  \vec{P}\cdot \vec{\Omega} \rangle$ to form regardless of the angle between the star's rotation and magnetic field.  A neutron star spins on the order of milliseconds and the kick occurs over hundreds of years.  This  gives rise to a cylindrical symmetry which causes the the force of the kick to average along the axis of rotation.  The nature of the spin-magnetic field correlation can produce kicks that are either aligned or anti-aligned with the spin axis. Regardless of its direction, this correlation is a P-odd effect, which indicates it must be generated by a P-odd mechanism such as the parity violating processes that power our current.  In fact, long, sustained kicks are observationally supported by the analysis in \cite{Spruit:1998} and the spin-kick correlation is observationally supported in \cite{Spruit:1998,Wang:2006zia,Rankin:2007}.  This can be considered indirect  support for our proposal.
 
Our mechanism is similar to neutrino emission in that we are ejecting a particle from the star, but there are some key differences. Neutrino emission is automatically  asymmetric with respect to the direction of the magnetic field $\vec{B}$; however, most neutrino-based kick mechanisms have difficulty delivering the produced asymmetry to the surface of the star.  Only when it can reach the surface can the asymmetry push the star.   In most neutrino  based mechanisms the star must be very hot ($T >$ MeV) for neutrinos to be energetic enough to transfer sufficient momentum and the kick occurs in a matter of seconds. But at such high temperatures the neutrinos cannot escape the star without interacting and washing out the asymmetry. 

We also suggest that a current running along the magnetic flux (the poloidal field) may be the source of the toroidal magnetic field and the finite magnetic helicity thought to be required in a neutron star \cite{Wright:1973,Markey:1973,Flowers:1977,Spruit:2007bt}.  The magnetic helicity, $\mathcal{H}\equiv \int d^3x \vec{A}\cdot\vec{B}$, that arises from the linking of toroidal and poloidal magnetic fields is a topological invariant and is a P-odd effect that must be generated though parity violation.   Our method of generating a toroidal field naturally produces magnetic helicity without requiring arguments to temporarily break the topological invariance of the helicity.  Many attempts at generating magnetic helicity rely on instabilities in the magnetic field caused by the star's rotation.  Such correlations $\langle \vec{B}\cdot\vec{\Omega}\rangle$ are P-even, and though they may generate toroidal fields, they cannot be responsible for helicity. 

These applications of topological currents to seemingly unrelated P-odd effects will be discussed in greater detail in Section 8 and only comprise a sample of possible applications.  Remarkably,  the effects of topological currents introduced in this paper  can be  experimentally tested in terrestrial laboratories, particularly in some condensed matter systems and in RHIC at Brookhaven.  
 
\section{A Big Picture: The Basic Ingredients and Assumptions} 
Topological vector currents can couple to charged fermions and become physical, superconducting, electromagnetic currents.  Our motivation comes from the possible effects of topological currents on the  physics of neutron stars.  Even though axial currents exist in the star they cannot be used as an electromagnetic source and we will not consider them.  Non-dissipating, induced vector currents at $T=0$ have the form   \cite{Alekseev:1998ds,Charbonneau:2007db,Fukushima:2008xe} 
\begin{eqnarray}  
\label{j}
\langle j \rangle = (\mu_l - \mu_r)\frac{e \Phi}{2\pi^2},
\end{eqnarray}  
where $\mu_r$ and $\mu_l$ are the chemical potential of the right and left-handed electrons, and $\Phi$ is the magnetic flux.   For the current to be nonzero the number of left-handed particles must be different than the number of right-handed particles.  The weak interaction, which strongly violates parity, is a natural source for the required asymmetry and is where we focus our attention.

We will assume that electrons are the only reasonable charge carrier because all other charged particles in the star are too heavy.  The particles in a neutron star attain equilibrium through the weak interaction, which creates predominantly left-handed particles.  This creates an intrinsic difference in the number of left-handed and right-handed electrons.  In an infinite system this imbalance would disappear---the average helicity of the electrons would be washed out due to the inverse weak P-violating processes.  The key is that the neutron star is a finite system and electrons are removed from the star by the current before they can decay.  The asymmetry that created the current is allowed to propagate to the surface and not get washed out.\footnote{Here and in what follows we neglect all QED re-scattering effects, which are much stronger than weak interactions but they are P-even and, therefore,  cannot wash out the produced asymmetry.  This is discussed in detail in Section 3.2.  Because of the large magnetic field the electron only travels in the direction of the magnetic field while the motion in transverse directions confined to Landau levels. The  term ``mean free path'' in this paper implies  the weakly interacting P-odd  ``mean free path'" when a produced asymmetry can be washed out.  } 

This topological current corresponds to the lowest energy state in the thermodynamic equilibrium when $ \mu_l \neq \mu_r$ is held fixed. In reality there is a tendency for $ \mu_l $ and $\mu_r$ to equilibrate though weak interactions; however, due to the finite size of the system a complete equilibrium cannot be achieved.  This is analogous to how neutrinos in cold stars can leave the system without further interactions, but unlike with neutrinos the electron chemical potential does not drop to zero. The induced current only remains  non-dissipating when the system is degenerate, $\mu\gg T$.  In the star's crust this condition becomes invalid, the current will become dissipative, and the trapped electrons will return into the system.  A detailed discussion of this region when $\mu\sim T$ is beyond  the scope of the present work, and will be discussed somewhere else.  If the electrons manage to escape the star the electron chemical potential will slowly decrease until the current may stop running.  Charge neutrality will cause matter to accrete isotropically possibly maintaining some of the chemical potential. 

In the following subsections we will discuss the structure and processes of dense matter, formally derive the topological current, and discuss in greater detail how the magnitude of the current can be estimated.

\subsection{Notation}
We use the convention $\hbar = c = k_B = 1$ unless otherwise stated.  We will always denote the momentum of particle as $p_i$, the Fermi momentum as $k_i$, and the chemical potential as $\mu_i$. When convenience dictates, the subscript $i$ will either be the symbol of the particle or a number, which will be labelled on the Feynman diagram.  The three momentum will be bolded $\bm{p}_i$, with a magnitude denoted $p_i$, and the four momentum will have a greek index $p^\mu$.

\subsection{Topological vector currents}
The purpose of this section is to provide a brief introduction to topological currents and explicitly derive the form of the vector current.   We will only provide a sketch here and refer the reader to \cite{Metlitski:2005pr,Metlitski:2005qz,Fukushima:2008xe}  for the finer details.  

Consider the vector current $j_{\textrm{v}}^3 = \bar{\psi}\gamma^3\psi$ in the presence of a magnetic field pointing z-direction, where $\psi$ is a Dirac spinor.  Ultimately we are interested in the left and right-handed modes and write the current in the Weyl representation,
\begin{equation}
j_{\textrm{v}}^3 = \psi_l^\dagger \sigma^3 \psi_l - \psi_r^\dagger \sigma^3 \psi_r\,.  
\end{equation}
The Dirac equation in a magnetic field with potential $\mathbf{A}$ can be written as 
\begin{eqnarray}
(H^2 + m^2)\psi_r  = E^2\psi_r,
\end{eqnarray}
where  $\psi_l = \frac{1}{m}(E-H)\psi_r$.  Eigenvalues of $H$ acting on $\psi_r$ are labelled $\epsilon$, hence the Dirac equation has two solutions, $E_\pm=\pm\sqrt{\epsilon^2+m^2}$.  The Dirac spinor can be written entirely in terms of the right-handed spinor and its eigenvalues,
\begin{eqnarray}
\psi_{\pm} = \begin{pmatrix} \psi_l \\ \psi_r \end{pmatrix}_\pm = \frac{1}{[4(m^2+\epsilon^2)]^{1/4}} \begin{pmatrix} \pm[(m^2+\epsilon^2)^{1/2}\mp \epsilon]^{1/2} \\ [(m^2+\epsilon^2)^{1/2}\pm \epsilon]^{1/2} \end{pmatrix} \psi_r
\end{eqnarray}
We further break up the operator $H$ into transverse and longitudinal components $H = p_3 \sigma^3 + H_\perp$, where $p_3$ are momentum eigenstates along the magnetic field and $H_\perp=(-i\partial_i + eA_i) \sigma_i$ is a transverse operator with eigenstates $|\lambda\rangle$.  We take the longitudinal direction to be periodic in $L$ and take the limit $L \rightarrow \infty$ later.  One can prove that the zero modes of $H_\perp$ are simultaneously eigenstates of $H$ with eigenvalue $\epsilon=p_3\sigma^3$.

The expectation value for the current is found in the usual manner by summing the current over all states weighted by the probability of each state.  For fermions the probability is given by the Fermi-Dirac distribution.  The derivation for the axial current assumes that the densities for the left and right-handed modes are equal.  Here, in the derivation of the vector current, we consider the possibility that the densities are different and assign the left and right-handed modes each their own Dirac distribution, $n_l$ and $n_r$, in the expectation value,   
\begin{eqnarray}
\label{definition} 
\langle j_{\textrm{v}}^3\rangle &=& \sum_E \left[n_r(E)\ \psi_r^\dagger \sigma^3 \psi_r - n_l(E)\ \psi_l^\dagger \sigma^3 \psi_l \right]  \nonumber\\
&=& \sum_\epsilon \left[ n_r(E_+) + n_r(E_-) - n_l(E_+) - n_l(E_-) \right]\psi_r^\dagger\sigma^3\psi_r \,,
\end{eqnarray}
where we used the spinor definition above and the fact that summing over terms odd in $\epsilon$ vanish.  If we write everything in terms of the eigenstates of $H_\perp$ it can been shown that only the zero modes of $\lambda$ survive,
\begin{equation}
\langle j_{\textrm{v}}^3\rangle = \frac{1}{L} \sum_{p_3} \sum_{\lambda = 0}  \left[ n_r(E_+) + n_r(E_-) - n_l(E_+) - n_l(E_-) \right] \langle\lambda |\sigma^3|\lambda\rangle
\end{equation}

Following the standard arguments, the factor $ \langle\lambda |\sigma^3|\lambda\rangle$ counts the difference in transverse zero modes travelling parallel to the magnetic field with positive or negative eigenvalues, $N_+$ and $N_-$.  Taking $L \rightarrow \infty$ and integrating each dirac distribution gives the number density of 1-dimensional left and right-handed fermions, $n(T,\mu)_r - n(T,\mu)_l$. With this all taken into account the current can be written as, 
\begin{equation}
\label{1d}
\langle j_{\textrm{v}}^3\rangle  = [n_l(T,\mu) - n_r(T,\mu)](N_+ - N_-).
\end{equation}
The difference in the positive and negative modes travelling along the magnetic field is given in physical terms by the index theorem \cite{Metlitski:2005pr}, 
\begin{equation}
N_+ - N_- = \frac{e\Phi}{2\pi}\,, 
\end{equation}
where $\Phi$ is the magnetic flux.  The current is then,
\begin{equation}
\label{eq:current}
\langle j_{\textrm{v}}^3 \rangle = (n_l - n_r)\frac{e\Phi}{2\pi}\,,
\end{equation}
where $n_r(T,\mu)$ and $n_l(T,\mu)$ are one dimensional number densities of left and right-handed Dirac fermions. Furthermore, the one-dimensional number density in the massless limit can be written in terms of the chemical potential $n(T,\mu) = \mu/\pi$ in which case the expression for the current is reduced to (\ref{j}). In this case it is purely topological.  At $T=0$ the density one dimension density of massive particles has simple expression $n(0,\mu) = \sqrt{\mu^2-m^2}/\pi$. The mass slightly diminishes the magnitude of the current.  

\begin{figure}[ht]  \begin{center}
  \includegraphics[width=8cm]{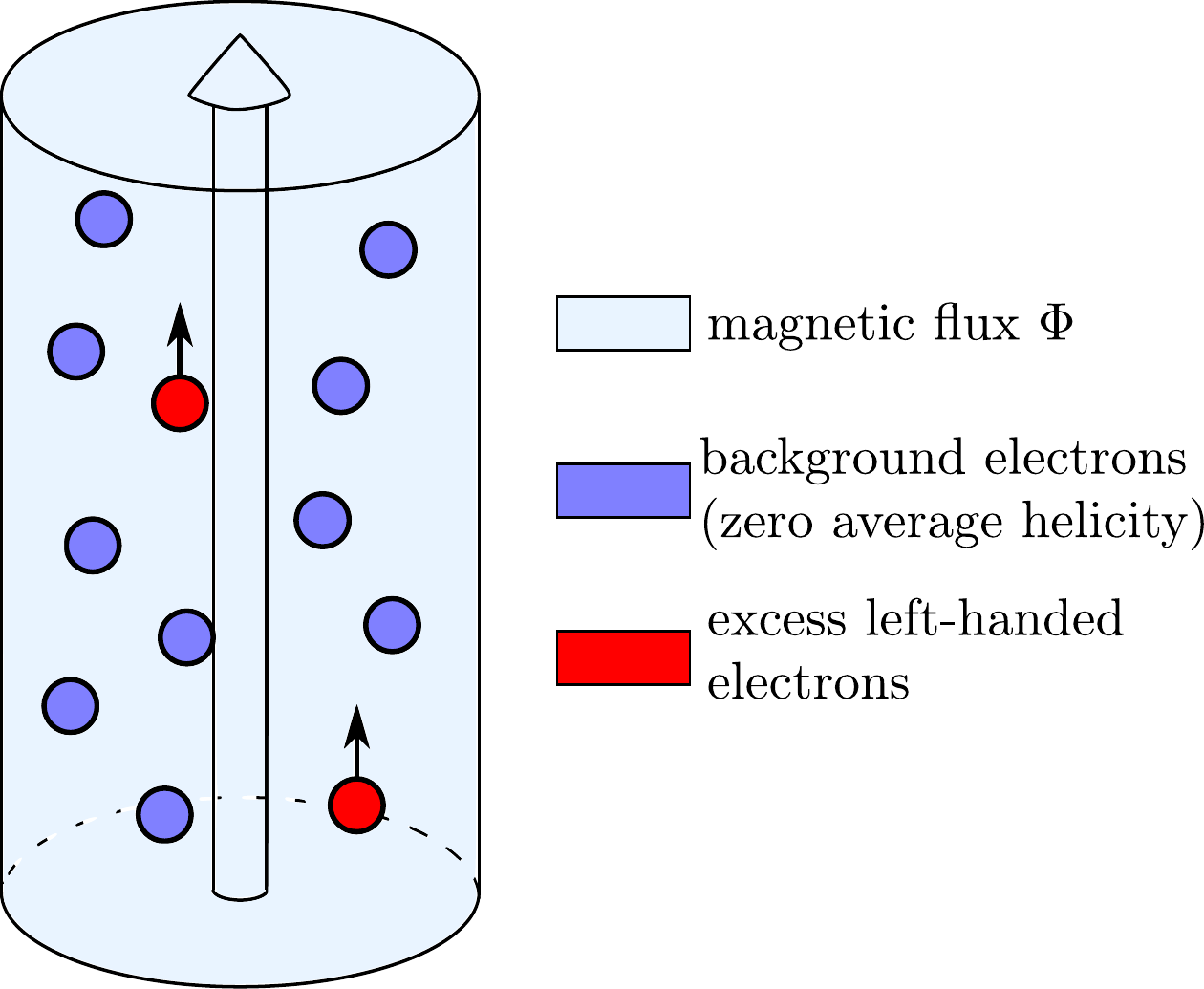}
  \caption{A left-handed electron placed inside the magnetic flux in a background of electrons with no average helicity will be pushed out.  The current wants to balance the number of left-handed and right-handed particles and acts as a helicity pump.}
  \label{fig:figure_flux}
  \end{center}
\end{figure}

These currents are not exotic, but are simply statements of the motion of left and right-handed particles given their spin alignment in a magnetic field.  Formulae (\ref{definition}) and (\ref{eq:current}) have a very simple physical meaning: to compute the current one should simply count the difference between left-handed and right-handed modes in the background of a magnetic field, see Figure \ref{fig:figure_flux}.  We assume that the modes of the current couple to electrons.  In this formulation we only consider the lowest landau level, which means the spin of an electron will antialign with the field.  Then if there is an excess of left-handed electrons they must move in the direction of the magnetic field and if there is an excess of right-handed electrons they must move against the field.  The topological vector current acts as a pump to remove the average helicity and it stops pumping once the average helicity is zero again.   

In general the difference in left and right-handed modes is a complicated function of many parameters: magnetic field $B$, chemical  potential $\mu$, temperature $T$, and mass of the particle $m$; but in the chiral limit ($m=0$) the expression for the current  takes  the simple  form (\ref{j}), has pure topological character, and can be derived from an anomalous effective Lagrangian without referring to the dynamics. The current is expressed  in terms of one-dimensional fermi distributions (\ref{1d}) and the physics of  two other dimensions is determined by Landau levels (the lowest one for $m=T=0$). When $T\neq 0$ and  $m\neq 0 $ the current will  still be induced, but it  will  not have a simple topological form (\ref{j}).  The relevant  formula in this case becomes much more complicated and is determined by the  ratios  of  a number of dimensional parameters mentioned above, see \cite{Fukushima:2008xe} where some limiting cases have been studied.

The current is insensitive to the structure of the magnetic field and will  be induced if the magnetic field is confined to magnetic flux tubes  or uniformly distributed.  The current is simply confined to the regions where the magnetic field is present.  For our purposes it is not essential whether the magnetic field is represented by magnetic flux tubes, as found in type-II superconductors, or by magnetic domains, the typical structure for  the intermediate state\footnote{The intermediate state is  characterized by alternating domains of superconducting and normal matter where the superconducting domains exhibit the Meissner effect, while the normal domains carry the required magnetic flux.}  argued for in argued for in \cite{Charbonneau:2007db}. The current is strongest in the degenerate regions with $\mu\gg T$  where the background magnetic field is large and when the system becomes a less degenerate (or not degenerate at all) one should expect strong suppression \cite{Metlitski:2005pr,Fukushima:2008xe}.   For numerical estimates of the effect it is convenient to count number of superconducting flux tubes  in the entire star and compute the current per unit quantum  flux  $\Phi_0=\frac{2\pi}{q}=2\cdot 10^{-7}$~G$\,$cm$^2$, where  $q=2e$ is the charge of the proton Cooper pair. 

Finally, we note that two different  applications of the induced vector current have been previously  discussed: a) neutron stars \cite{Charbonneau:2007db} and b)   RHIC physics  \cite{Fukushima:2008xe}. Our basic objective remains the same as in our previous paper  \cite{Charbonneau:2007db}; however the main focus in this paper will be estimating the magnitude of the induced current itself while our previous paper \cite{Charbonneau:2007db} was mainly devoted to the specific application on study of the type-I vs type-II superconductivity as a result of induced vector current.

\subsection{Basic weak processes }
The primary composition of nuclear matter in a neutron star is constantly being debated.  Fundamentally a neutron star is made of neutrons with small, equal fractions of protons and electrons.  In more exotic models hyperons may appear along with pion and kaon condensates.  In an effort to simplify the discussion we will constrain ourselves to four fundamental interactions that describe the majority of cases in dense stars. 

These interactions have been discussed before in the context of cooling a neutron star through neutrino emission.  The first two are quite closely related---the direct and modified Urca processes,
\begin{eqnarray}
\label{eq:Urca}e^- + P &\rightarrow& N + \nu_e \\
e^- + P  + N^\prime &\rightarrow&N^\prime + N + \nu_e \,,
\end{eqnarray}
which both obey the beta equilibrium condition $\mu_e + \mu_P = \mu_N$.  The neutrinos created in these processes only interact with matter through the weak force, which is so weak that the star is transparent to the neutrinos.  They leave the star and do not contribute to the equilibrium condition.   

The first of these interactions, the direct Urca Process, should be the dominant process in normal nuclear matter but it is heavily suppressed because the the particles are unable to conserve momentum while remaining on their Fermi surfaces.  It is possible for the direct process to conserve momentum if the proton fraction in the star is above 1/9 \cite{Lattimer:1991ib}, which could occur with the appearance of hyperons \cite{Page:2006ud}.  In these cases the direct Urca process dominates.    The modified Urca process is able to conserve momentum through an external nucleon and was used in the first neutrino emission calculations \cite{Bahcall:1965zzb}.  This is the dominant electron producing process in normal nuclear matter, but it is very slow and the presence of exotic particles introduces processes that create and destroy electrons quicker.  

As the density of matter increases it is likely that kaon  \cite{Kaplan:1986yq} and pion condensates will appear. We will restrict the discussion to kaon condensates, which appear at much more reasonable densities, $3\,n_0$, than pion condensates, $300\,n_0$, but the phenomenology of dealing with the two condensates is almost identical.  Electrons are still created and destroyed by the previous interactions, but at a much slower rate than an electron decaying\footnote{Here and in what follows the term ``electron decay" means the  transforming of  an electron  into a neutrino as a result of interactions with surrounding hadrons.}  into a kaon and neutrino in the presence of nucleons,
\begin{eqnarray}  
\label{eq:kaon}
e^- + N &\rightarrow& \langle K\rangle^- + N + \nu_e. 
\end{eqnarray}
This interaction and its inverse process add another equilibrium condition, $\mu_e = \mu_K$, on top of the previously mentioned beta equilibrium.  

The previous three interactions encompass the creation of electrons in almost all possible neutron star interiors.  The last interaction we consider is the primary source of electrons in quark stars.  The direct Urca process for quarks are,
\begin{eqnarray}  
\label{eq:quark}
e^- + u &\rightarrow& d + \nu_e \\ 
e^- + u &\rightarrow& s + \nu_e\,.
\end{eqnarray}
Unlike in normal nuclear matter there is no trouble conserving momentum in quark matter.  The direct process occurs unsuppressed and there is no need to discuss a modified process \cite{Iwamoto:1982}.

\subsection{Simple models for dense matter}
We will review the features of neutron matter that are required for the rest of the paper.  Most importantly we will summarize the short reviews found in \cite{Iwamoto:1982,Maxwell:1977zz,Bahcall:1965zza,Brown:1988ik} and state the values used in the rest of the paper. The simplest model is the non-interacting gas model where the ground state of the neutron star, $T=0$, is a mixture of neutrons, protons, and electrons that is electrically neutral.  The baryon density is on the order of nuclear density $n_0 = 0.17 $ fm$^{-3}$, which leads to the nucleons and electrons being highly degenerate.  

The particles achieve equilibrium though the Urca processes \eqref{eq:Urca}.  The neutrinos produced in these reactions only react weakly in the star and easily leave it at low temperatures.  Because of this neutrinos are often assumed to be non-degenerate and the chemical potentials in the star satisfy,
\begin{eqnarray} 
\mu_e + \mu_P = \mu_N\,.
\end{eqnarray}
Charge neutrality implies that $n_e = n_P$, which implies that the Fermi momenta of the electrons and protons are equal, $k_e = k_P$.  This restriction has two important effects.  The electrons are relativistic and the protons are non-relativistic implying that the chemical potential of the proton is much smaller that of the electron.  This further implies that the density of neutrons is much higher than the electrons and protons, thus a neutron star.  Assuming that the density of the neutrons is that of nuclear matter then,
\begin{eqnarray}    
k_N = (3\pi^2 n_N)^{1/3} \approx 340\,(n/n_0)^{1/3} \textrm{ MeV.}
\end{eqnarray} 
Equating the electron and neutron chemical potentials yields,
\begin{eqnarray}    
k_e \approx \frac{k_N^2}{2m_N}  \approx 62\,(n/n_0)^{2/3} \textrm{ MeV.}
\end{eqnarray}
In reality there is a correction to the non-interacting model due to the proton being more bound than the electron.  It is common in literature to assume a value of
\begin{eqnarray}    
k_e \approx 100\,(n/n_0)^{2/3} \textrm{ MeV,}
\end{eqnarray} 
which is the value we will use through out the paper.  

As the density of the star raises above $3n_0$ there is the possibility that $K^-$ condensates will appear \cite{Kaplan:1986yq}.  As the density of the nuclear matter increases, the density of the electrons increases to a point where it becomes advantageous to decay into negatively charged kaons though the process \eqref{eq:kaon}.  The system reaches equilibrium through the inverse process.  These processes add an additional equilibrium condition,   
\begin{eqnarray}    
\mu_e = \mu_K\,.
\end{eqnarray}
Though negatively charged pions are lighter than kaons, they are unlikely to appear until much higher densities due to the strong interaction increasing the effective mass \cite{Page:2006ud}.  

At around the same density as kaons appear, light hyperons and muons may appear \cite{Page:2006ud}.  The existence of hadrons lowers the neutron density, which lowers the ratio of neutrons to protons, making it possible for the direct Urca process to proceed unsuppressed, and opens up processes such as $\Lambda \rightarrow e^- + P + \bar{\nu}_e$, which are also not kinematically suppressed.  These processes occur at about the same rate as the direct Urca processes, so we will take the direct Urca process as a reasonable substitute for them.  The rate can then be adjusted by an integer factor to compensate for additional processes.  The appearance of muons has no effect of our calculations as it is too heavy to couple to the current. 

When the density gets high enough it is possible that the quarks deconfine---the hadrons break down into their constituent quarks.  For this paper we will consider only the existence of light quarks in the star, which attain equilibrium through the quark Urca processes \ref{eq:quark}.  The equilibrium conditions are, 
\begin{eqnarray}    
\mu_u + \mu_e &=& \mu_d \\
\mu_u + \mu_e &=& \mu_s\,, 
\end{eqnarray}
where, as in earlier cases, the neutrinos are not trapped and are not degenerate.  The quark matter must also be electrically neutral,
\begin{eqnarray}    
Q/e = \frac{2}{3}n_u - \frac{1}{3}n_d - \frac{1}{3}n_s - n_e = 0\,.
\end{eqnarray}
The simplest models assume that the quark masses are all zero, thus their Fermi momenta are equal to their Fermi energies, and the predicted electron density is zero.  This, however, is not adequate as leptons do exist in the star.  Room for the electrons comes from the large mass of the strange quark.  Though the up and down quarks are relatively light, $m_u \sim m_d \sim 5-10$ MeV, the strange quark mass is actually quite large, $m_s \sim 100-300$ MeV. The strange quark is nonrelativistic meaning there will be fewer of them and electrons must be present to conserve charge.  For this paper will will follow \cite{Iwamoto:1982} by assuming that the quarks are massless, $k_u\sim k_s\sim k_d \sim k_q$, where  
\begin{eqnarray}     
\label{eq:k_quark}
k_q = (\pi^2 n_b)^{1/3} \sim 235 \left( \frac{n_b}{n_0}\right)^{1/3} \textrm{ MeV}\,,
\end{eqnarray}
where $n_b$ is the baryon number density.  For typical densities in the core of the neutron star the Fermi momentum is $k_q \sim 400$ MeV. We can approximate the electron Fermi momentum using the fraction of electrons to baryons, $Y_e = n_e/n_b$, which yields,
\begin{eqnarray}     
k_e = (3Y_e)^{1/3}k_q\,.
\end{eqnarray}
The typical value for the electron fraction is $Y_e = 0.01$.
 
We have discussed the nature of topological currents and determined the requirements for them to exist.  We have also given an overview of many types of dense stars, which are the only viable places for these current to appear.  Now we are ready to discuss the details of how the current appears in the star, and what its magnitude would be.

\subsection{Structure of the magnetic field in a neutron star}
The topological current is confined to regions of the neutron star where there is a magnetic field, but the magnetic flux structure inside a neutron star is non-trivial.  Directly calculating the current would require careful consideration of the helicity of electrons in regions with flux, where all electrons created are left-handed, and regions without flux, where the helicity can be  washed out, and how electrons diffuse from one region to the other.  In estimating the current in Section 3 we will simplify this by considering what would happen if the flux were uniformly and continuously spread throughout the star.  The magnitude of the current in the entire star depends only on the amount of flux, not its structure.  The total current leaving the star would be the same if some complex structure were present. Bunching flux lines would simply mean there are smaller regions with stronger magnetic fields.  However, it is important to be aware of this non-trivial structure as it will be considered later in Section 8 when we discuss applications of the current.

A typical neutron star has a large magnetic field, $B\sim10^{12}$ G.  Once it has cooled, the star has a temperature $T\sim 10^9 $ K that is cold compared to its Fermi energy $\mu\sim 100 ~\text{MeV} \sim 10^{12}$ K. Because of this the the protons are likely superconducting and the neutrons form a superfluid.  The magnetic field is large enough that it is favourable for the flux to penetrate the superconductor, rather than being completely expelled. The Meissner effect forces the flux to bundle into into either type-II vortices or type-I domains, which is often called the intermediate state.  It is generally believed that the protons form a type-II superconductor.  The Landau-Ginzburg parameter $\kappa=\lambda/\xi$ determines the type of superconductivity.  Typically in a neutron star the London penetration depth is $\lambda \sim 120$ fm and the coherence length of the proton superconductor is $\xi \sim 30$ fm.  This creates a ratio $\kappa > 1/\sqrt{2}$, which indicates type-II superconductivity.  For a type-II superconductor the magnetic field will penetrate the star by destroying narrow regions of superconductivity that each carry a single quantum of flux, $\Phi_0 = 2\pi/q=2\cdot10^{-7}$ G$\,$cm$^2$.  But there are problems with this picture \cite{Link:2003hq,Link:2006nc}.   It is possible that the system behaves as  type-I superconductor even though the Landau-Ginzburg parameter would suggest type-II behaviour \cite{Charbonneau:2007db}.   The argument in  \cite{Charbonneau:2007db} relies on the electromagnetic interaction between current carrying vortices rather than altering the value of the Landau-Ginzburg parameter $\kappa$.

If the intermediate state is realized in neutron stars the magnetic field distribution will be again non-uniform, but the structure would be quite different.  The intermediate state is characterized by alternating domains of superconducting and normal matter where the superconducting domains exhibit the Meissner effect, while the normal domains carry the required magnetic flux.  The pattern of these domains is strongly related to the geometry of the problem, see \cite{Charbonneau:2007db} for details.   While precise calculations are required for understanding of the magnetic structure in this case  one can give the following   estimation for  typical size of a domain  as suggested in \cite{Charbonneau:2007db,Buckley:2003zf}
\begin{equation}
\label{domain}
a\sim 10 \sqrt{R \lambda}\,,
\end{equation}
where $R$ is a typical external size identified with a neutron star core ($R\sim 10$ km), while $\lambda$ is a typical   microscopical scale of the problem. Numerically $a\sim 10^{-1}$ cm, which implies that a typical domain can accommodate about $10^4$ neutron vortices separated by a distance $10^{-3}$ cm. While the field distribution for the intermediate state  and the  type-II superconductor  are very different one should anticipate that the ratio of normal to superconducting regions are the same. 

Regardless of the flux structure, there are two regions of the neutron star to consider---those with magnetic flux and those without.  The total units of quantum flux can be estimated as 
\begin{eqnarray}    
 \label{N}
 N_{\textrm{v}}\sim \frac{\pi R^2 B}{\Phi_0}\sim 10^{31}B_{12}, \quad B_{12}\equiv \left(\frac{B}{10^{12}G}\right).
 \end{eqnarray}
The region that a single unit of flux occupies has a radius equal to the London penetration depth of the field $\lambda \sim 100$ fm.  This is multiplied by the number of vortices $N_{\textrm{v}}$ to get the total area.   If we take a slice of the neutron star perpendicular to the magnetic field we find that the ratio of the area occupied by flux tubes  is much smaller than the area occupied by the void,
\begin{eqnarray}    
\label{A}
\frac{A_{\textrm{vortices}}}{A_{\textrm{star}}} \simeq \frac{N_{\textrm{v}}\pi \lambda^2}{\pi R^2}
\simeq   \frac{\pi \lambda^2 B}{\Phi_0} \simeq 10^{-3}\cdot B_{12}\,.
\end{eqnarray}
This suppression essentially reflects the difference between typical magnetic field $B\sim 10^{12}$ G
and the critical magnetic field $B_{\textrm{c1}}\sim 10^{15} $ G when the superconductivity is destroyed. 

\section{Estimating Non-dissipating Currents in Dense Matter}

There are three requirements for topological vector currents to be present: an imbalance in the number of left and right-handed particles $n_l\neq n_r$, degenerate matter $\mu\gg T$, and the presence of the background magnetic field $B\neq 0$.  All of these  are present in neutron and quark stars.  The weak interactions, which the star attains equilibrium through, violate parity; particles created in this environment are primarily left-handed, see the Appendix for a quantitative estimates.  As discussed the interior of the star is dense, $\mu_e = 100$ MeV,  and cold, $T=0.1$ MeV, such that the degeneracy condition $\mu\gg T$ is met, and the star is known to have  a huge magnetic field, $B\sim10^{12}$ G.  

\begin{figure}[ht]  \begin{center}
  \includegraphics[width=8cm]{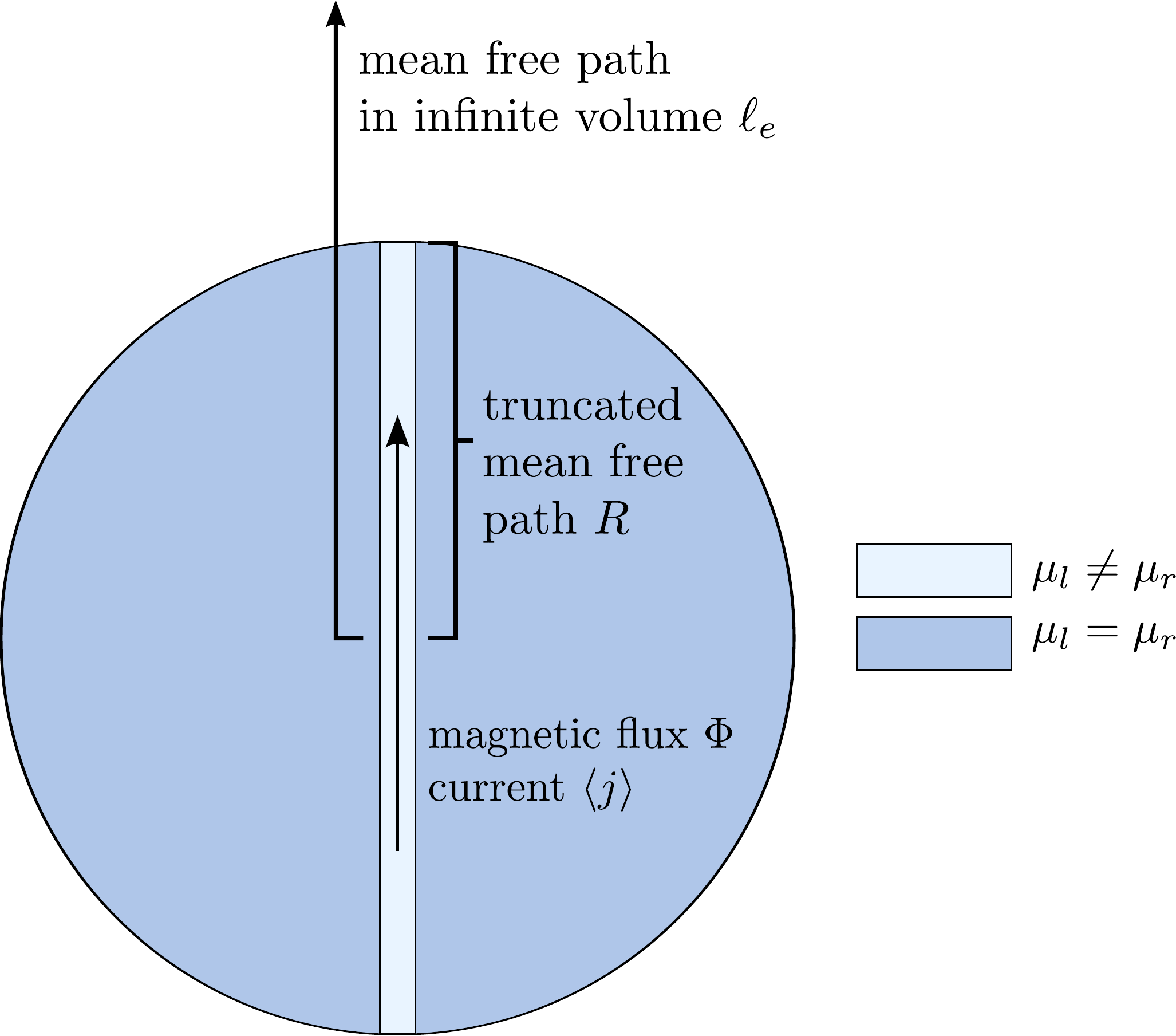}
  \caption{The neutron star is a finite system---the mean free path with respect to the weak interaction is much larger than the radius of the neutron star $\ell_{\textrm{weak}}\gg R$.  Electrons leave the system before they decay and contribute towards the current.  The current flows because the system is not in equilibrium.}
  \label{fig:figure_mfp}
  \end{center}
\end{figure}

All three criteria are met but there is a subtlety to consider.  In an infinite system, a system large enough to allow the electron to decay as discussed in section 2.3, any asymmetry in left and right-handed electrons created by the weak interaction would be washed out; the creation and annihilation rates of the left-handed particles are the same.  Though many more left-handed electrons are created, they are also destroyed such faster than the right-handed and no asymmetry builds. This is similar to the argument found in \cite{Kusenko:1998yy}---there is no asymmetry in equilibrium in an infinitely large system.  Unlike in \cite{Kusenko:1998yy}, we are interested in low temperatures where the neutron star is a finite system with respect to the weak interaction---see Figure \ref{fig:figure_mfp}.  The weak interaction can no longer remove the parity introduced into the system. As a result a current forms to remove it. We will first explain how the topological currents are induced. We will then discuss why the induced current is not  washed out by fast quantum electrodynamic (QED) processes, but can only be washed out by slow weak interactions.  The reason is the P-odd nature of the phenomenology: an effect induced by P-odd forces, can only be washed out by P-odd forces.

\subsection{How topological currents are induced}
If we assume that the magnetic field is uniformly distributed (as discussed in Section 2.5) then every electron created by a P-odd process in the star potentially contributes to the current.  As discussed in Section 2.2 we can calculate the current by counting the number of left-handed electrons minus the number of right-handed electrons created in the star.  Unlike Section 2.2 there is now a large Fermi momentum that opens up many Landau levels, but only the lowest Landau level contributes to the current. The current arises because the lowest Landau level has a single spin state, electrons created in that spin state are primarily left-handed, and this helicity state propagates out of the star, preserved through through countless QED interactions, before the inverse weak process can remove it.  These helicity states propagating out of the star make up the current.   

We explicitly see how both the magnetic field and parity violation are necessary for the current. The spin degeneracy of the lowest Landau level is one, while all other Landau levels have spin degeneracy two.  This implies that the produced P-asymmetry in the polarization  $ \langle \Lambda \rangle =  \langle \vec{\sigma}\cdot \vec{P}/| \vec{P}|\rangle$ is not translated into P-asymmetry in the correlation of the momentum and magnetic field $ \langle\vec{B}\cdot \vec{P}\rangle$ for all Landau levels except the lowest one. Even though the P-asymmetry is present in the higher levels the spin degeneracy allows particles with the same polarization  are allowed to travel in opposite directions, which results in zero current. The single spin degeneracy of the lowest Landau level means that any longitudinal polarization will result in more particles moving one way than the other, thus a current.     The correlation between spin and the magnetic field is a P-even effect $ \langle \vec{\sigma}\cdot \vec{B}\rangle$ and together with the P-odd correlation between spin and momentum $\langle \vec{\sigma}\cdot \vec{P}\rangle$ it produces the P-odd asymmetry $ \langle \vec{B}\cdot \vec{P}\rangle$ we are interested in.  

Equation (\ref{eq:current}) tells us how to proceed. The current is found by counting the difference in the creation rate of the left and right-handed electrons created in the lowest Landau level. This results in a current of the form,
\begin{eqnarray}
\label{current-new}
\langle j \rangle =   P_{\textrm{asym}}(B, \mu, T) \cdot  \frac{w}{\Omega}V_{\textrm{star}}\,, 
\end{eqnarray}
where $w/\Omega$ is the transition rate per unit volume assuming the magnetic field is uniformly distributed,  $ P_{\textrm{asym}}(B, \mu, T) $ accounts for both the  polarization  of electrons created in a magnetic field and the suppression due to Landau levels, and $V_{\textrm{star}}$ is the volume of the region of degeneracy, which we assume is about the same size as the star itself.  

The extreme degeneracy of the electrons means that electrons are uniformly distributed throughout the Landau levels.  As discussed in Section 4.2 the number of Landau levels is $n_{\textrm{max}} = (E_e^2 - m_e^2)B_{\textrm{c}}/2m_e^2B$ and the ratio of the lowest Landau level to the total number of states is  $\sim m_e^2 B /\mu_e^2B_{\textrm{c}} $, which is supported by the analysis in \cite{Sagert:2007as}. Therefore, we estimate  P-odd asymmetry factor as
\begin{equation}
\label{P}
P_{\textrm{asym}}(\mu_e, T, B)  \simeq  - \langle \Lambda \rangle\cdot  \left(\frac{m_e^2}{\mu_e^2}\frac{B}{B_{\textrm{c}}}\right) \sim 2\cdot10^{-5} \left(\frac{B}{B_{\textrm{c}}}\right)\left(\frac{n}{n_0}\right)^{-4/3},
\end{equation}
where the polarization $\langle \Lambda\rangle $ is calculated in the appendix and is  given by (\ref{helicity}).  Numerically we see that a large chemical potential means that the number of states in the lowest Landau level is small compared to the total number of states.  We have calculated the average polarization  not just for the lowest level, but for electron created in all levels.  Because we are looking at an average current in a large system it is likely that polarization  states other than those in the lowest landau level are the ones to propagate.  What is important is that there are more spin-down states than spin-up states and there is an average polarization.  The density dependence in (\ref{P}) reflects that found in a neutron star; in the case of quark stars we use the appropriate electron chemical potential and density dependence described in Section 2.4.   The suppression factor is identical to (\ref{b}) but appears because we are only considering the Landau levels that contribute to the current.  This is different than in (\ref{b}) where the factor appears because only one Landau level exists in the system.  

For non-trivial vortex structures it is convenient  to determine the current per unit fundamental flux by dividing by the total number of flux tubes.  In a type-II structure this would be the current that runs along a single vortex; in a uniformly distributed environment it is simply a convenient normalization, 
\begin{eqnarray}
\label{eq:current_estimate}
\langle j \rangle =  P_{\textrm{asym}}(B, \mu, T)\cdot \frac{w}{\Omega}\frac{V_{\textrm{star}}}{N_{\textrm{v}}}\,. 
\end{eqnarray}
The current in a type-I domain is found by multiplying (\ref{eq:current_estimate}) by the units of flux trapped inside the domain.

The expression for the induced current, (\ref{current-new}) and (\ref{eq:current_estimate}), naively has a different form than previously discussed (\ref{1d}), but in fact is precisely the same induced current with the same physical meaning.  The separation of transverse and longitudinal degrees of freedom  (with respect to magnetic field $B_z$) that is explicit in (\ref{1d}) is hidden now in the formula for $ P_{\textrm{asym}}(B, \mu, T)$ where the Landau levels  (transverse degrees of freedom) are treated separately from longitudinal motion. See the Appendix for details. The longitudinal degrees of freedom in eq.  (\ref{1d}) are represented  at $T=0$ by the one-dimensional number density $\sim \mu/\pi$, which is the correct expression when the problem is treated as a grand-canonical ensemble with  $\mu_L$ and $\mu_R$  constant  due to the infinitely large bath surrounding the system. In our case the neutron star is a finite system at $T\neq 0$ where particles are continuously injected at a rate $w$.  Along with $P_{\textrm{asym}}(B, \mu, T)$ this describes the resulting asymmetric number density as a function of external parameters $T, \mu, B$.  And as they should, both equations (\ref{1d}) and (\ref{current-new}) have units of current: number of particles per unit time. 

The topological vector current\footnote{The axial current will always be induced even when $\mu_l=\mu_r$, but it will  not be coupled to the electromagnetic field, and cannot carry the momentum. The physical consequences of this axial topological current might be quite interesting, but shall not be discussed in the present paper.}   arises specifically because the system is no longer in equilibrium with respect to the weak interaction and a small asymmetry $(\mu_L-\mu_R)\neq 0$ appears.  The current is a steady state\footnote{ We assume that the variation  of $(\mu_L-\mu_R)$ is adiabatically slow process,   with a typical variation time to be much longer than any other time scales of the problem. It allows us to treat the system   as being in the  equilibrium when $(\mu_L-\mu_R)$ is assumed to be a fixed parameter. }, but constantly requires new polarized electrons to be created in the magnetic field background, then pushed out of the system, into the crust or into space.  Because of this steady state the rate $w/\Omega$ is calculated when the electron chemical potential is constant.  

The electron chemical potential $\mu_e$ does not necessarily go to zero as it does for neutrino emission. Only a small fraction of electrons equal to the number in the lowest Landau level actually participate in the current.  When electrons leave the region of degeneracy, where the non-dissipating current is produced, the current loses its quantum properties and becomes a normal dissipating current capable of transferring momentum, emitting photons, etc. These electrons may get trapped in the crust and diffuse back in, maintaining $\mu_e$.  If the electrons manage to escape, the chemical potential   $\mu_e$ will slowly decrease over time.  As electrons escape the star becomes positively charged and stars to accrete matter, which may diffuse back into the star.  Eventually, when $\mu_e$ becomes sufficiently small, the production  of the  induced current stops. This is precisely the moment when the neutron star kick engine stops. As we mentioned previously it is the chemical potential that fuels the kick, not the temperature. The engine stops when chemical equilibrium is achieved, not when the temperature drops which happens much earlier.  

\subsection{P-odd effects, QED, polarization, and thermodynamics}

An important aspect of the topological current is that it is neither a zero temperature effect nor a chiral effect. It persists as long as parity, an asymmetry in left and right-handed electrons, is present in the system. Indeed, while the topological currents originally were computed at $T=0$ for exactly massless fermions~\cite{Son:2004tq}, it was later shown~\cite{Metlitski:2005pr} that the effect persists for $T\neq 0, m_e\neq 0$.   The important lesson learned from these calculations is that the effect is not washed out, even when a nonzero temperature is introduced and     an arbitrarily large number of collisions between electrons occur to maintain thermodynamic equilibrium. The crucial point for our mechanism is that P-odd effects are not washed out by fast QED processes.

It is important to remember  there are  three scales in our problem: the mean free path of quantum electrodynamic processes $\ell_{\textrm{QED}}$, the mean free path of weak interactions $\ell_{\textrm{weak}}$, and the radius of the neutron star $R$.  For the temperatures we are interested the scales are ordered as $\ell_{\textrm{weak}}\gg R \gg \ell_{\textrm{QED}}$.  We see that with respect to QED interactions the neutron star is considered to be an infinite system, but with respect to weak interactions it is finite. 

Quantum electrodynamic processes cannot wash out polarization even though massive particles introduce processes that can flip helicity, even when these processes occur thermally.  The key here is the unitarity of interactions; processes go forwards and backwards at the same rate.  Though helicity is lost through these interactions, it is also created.  This cancellation due to unitarity is aptly  illustrated in \cite{Kusenko:1998yy}, where the authors  consider the weak interactions inside a star at high temperatures.  Just as in \cite{Kusenko:1998yy} we can use thermal equilibrium and unitarity to show how the current propagates.

Consider electron-proton scattering where neither particle is confined to Landau levels.  The change in the number of electrons with a given helicity $n^{(e)}(\bold{q}\cdot\bm{\sigma},t)$  can be written as,
\begin{eqnarray}
\label{eq:boltzmann}
\frac{\partial}{dt}n^{(e)}(\bold{q}\cdot\bm{\sigma},t) &=& \sum_{\bm{\sigma},\bm{\sigma}^\prime,\bold{s},\bold{s}^\prime} \int \frac{d^3\bold{p}}{p_0}\frac{d^3\bold{p}^\prime}{p^\prime_0}\frac{d^3\bold{q}^\prime}{q^\prime_0} \\ && \nonumber \left[  \mathcal{S}\ W(\bold{q}\cdot\bm{\sigma},\bold{p}\cdot\bold{s}|\bold{q}^\prime\cdot\bm{\sigma}^\prime,\bold{p}^\prime\cdot\bold{s}^\prime)  - \mathcal{S}^\prime\ W(\bold{q}^\prime\cdot\bm{\sigma}^\prime,\bold{p}^\prime\cdot\bold{s}^\prime|\bold{q}\cdot\bm{\sigma},\bold{p}\cdot\bold{s}) \right], 
\end{eqnarray}   
where $\bm{q}$ and $\bm{p}$ denote the momentum of the electron and proton, $\bm{\sigma}$ and $\bold{s}$ denote the electron and proton spin, and $W(\bold{q}^\prime\cdot\bm{\sigma}^\prime,\bold{p}^\prime\cdot\bold{s}^\prime|\bold{q}\cdot\bm{\sigma},\bold{p}\cdot\bold{s}) $ is the probability of scattering $| p(\bold{q},\bold{s})e(\bold{q},\bm{\sigma}) \rangle \rightarrow |  p(\bold{q}^\prime,\bold{s}^\prime)e(\bold{q}^\prime,\bm{\sigma}^\prime) \rangle$ per unit time per unit phase volume. These are all the processes that change the helicity of the electron.

There is a statistical factor $\mathcal{S}$ to account for  the Fermi distribution of the initial states and the Pauli blocking of the final particles. In the massive limit the chemical potential enters the Lagrangian by a term $\mu N$, where the particle number $N$ is the charge associated with the chemical potential. The mass term literally mixes up the left and right spinors in the Lagrangian. Because of this only one value can enter the statistics of the the problem. But the chemical potential can be written as $\mu_e = \mu_l + \mu_r$, where the left and right-handed chemical potentials are really the contribution towards the chemical potential rather than chemical potentials in their own right. The current is given by equation (\ref{eq:current}) where we calculate the difference in the numbers of left and right-handed electrons. It is common to work in the chiral limit where there is no mass term and it is possible in the Weyl representation to introduce two separate chemical potentials, one for the left-handed and one for the right-handed spinors.  There are then two charges that enter the statistics and there are two separate Fermi surfaces. The current could then be determined by looking at the difference in these chemical potentials. But in the chiral limit there are no helicity flipping amplitudes and the imbalance in left and right-handed chemical potentials created by the weak interaction would never be washed out.  The purpose of this section is to show that the current does not get washed out when helicity flipping amplitudes exist so we choose the massive limit where the electrons feel only a single chemical potential regardless of their helicity.  

With this in mind the statistical factor written explicitly is,
\begin{eqnarray}
 \mathcal{S} = \frac{1}{1+e^{-(E(p_e)-\mu_e)/T}} \frac{1}{1+e^{-(E(p_p)-\mu_p)/T}} \frac{1}{1+e^{(E(p_e^\prime)-\mu_e^\prime)/T}} \frac{1}{1+e^{(E(p_p^\prime)-\mu_p^\prime)/T}}\,.
\end{eqnarray} 
The factor $\mathcal{S}^\prime$ is similar but has the primed and unprimed variables swapped. We have chosen degenerate functions to illustrate that the average polarization of the electrons  along a specific direction (in our case along the magnetic field) is constant, and cannot be washed out by QED interactions, see eq. (\ref{constant}) below. We have made no assumptions about the initial and final chemical potentials and assume that they are different. 

The general statement of thermal equilibrium is 
\begin{eqnarray}
\label{eq:thermal_eq}
\mathcal{S} = \mathcal{S}^\prime\,,
\end{eqnarray} 
Thermal equilibrium is attained with respect to QED interactions because the mean free path is much smaller than the size of the neutron star.  This in not the case with the weak interactions, which have a mean free path much larger than that of the star.  This non-equilibrium of the weak interaction generates the parity asymmetry required for the current and is  discussed in detail in Section 3.1.  Given our specific distributions, the statement of thermal equilibrium has two solutions,
\begin{eqnarray}
E(p_e^\prime)-\mu_e^\prime = E(p_e)-\mu_e \quad \textrm{     and     } \quad E(p_p^\prime)-\mu_p^\prime = E(p_p)-\mu_p \,,
\end{eqnarray} 
or
\begin{eqnarray}
E(p_e)-\mu_e = E(p_p^\prime)-\mu_p^\prime \quad \textrm{     and     } \quad E(p_e^\prime)-\mu_e^\prime = E(p_p)-\mu_p\,. 
\end{eqnarray} 
Substituting either of these solutions into the conservation of energy equation for the interaction $E(p_e) + E(p_p) = E(p_e^\prime) + E(p_p^\prime)$ yields,
\begin{eqnarray}
\mu_e + \mu_p = \mu_e^\prime + \mu_p^\prime\,,
\end{eqnarray}
which is the condition for chemical equilibrium. Particles can only scatter from their Fermi surface onto another Fermi surface.

The key element of the argument is the unitarity of the interaction,
\begin{eqnarray}
\label{eq:unitarity}
1 &=& \sum_{\bm{\sigma}^\prime,\bold{s}^\prime} \int \frac{d^3\bold{p}^\prime}{p^\prime_0}\frac{d^3\bold{q}^\prime}{q^\prime_0} W(\bold{q}\cdot\bm{\sigma},\bold{p}\cdot\bold{s}|\bold{q}^\prime\cdot\bm{\sigma}^\prime,\bold{p}^\prime\cdot\bold{s}^\prime)\\ &=& \sum_{\bm{\sigma}^\prime,\bold{s}^\prime} \int \frac{d^3\bold{p}^\prime}{p^\prime_0}\frac{d^3\bold{q}^\prime}{q^\prime_0} W(\bold{q}^\prime\cdot\bm{\sigma}^\prime,\bold{p}^\prime\cdot\bold{s}^\prime|\bold{q}\cdot\bm{\sigma},\bold{p}\cdot\bold{s}),
\end{eqnarray}
This says that every initial state scatters into a final state and that every final state scatters into an initial state. 

Using (\ref{eq:thermal_eq}) the electron and proton distributions can be factored out and using (\ref{eq:unitarity}) we find that the forward and reverse rates cancel each other, causing the right hand side of (\ref{eq:boltzmann}) to vanish.  We have found that the number of electrons of a given helicity is constant, 
\begin{eqnarray}
\frac{\partial}{dt}n^{(e)}(\bold{q}\cdot\bm{\sigma},t)=0\,. 
\end{eqnarray}
This is the result of detailed balance, which ensures that there can be no asymmetry in the creation and annihilation rates of particles in thermal equilibrium.  Given this spectrum of static solutions, the average polarization of the electrons  along a specific direction (in our case along the magnetic field) is constant,
\begin{eqnarray}
\label{constant}
\langle \Lambda \rangle = \frac{\int d^3\bold{q} \left[ n^{(e)}(+|\bold{q}\cdot\bm{\sigma}|) - n^{(e)}(-|\bold{q}\cdot\bm{\sigma}|)\right]}{\int d^3\bold{q} \left[ n^{(e)}(+|\bold{q}\cdot\bm{\sigma}|) + n^{(e)}(-|\bold{q}\cdot\bm{\sigma}|)\right]} =\textrm{constant}.
\end{eqnarray}
The physical meaning of equation (\ref{constant}) is very simple:  if we start with no average  polarization, none develops; if we start with an averall polarization, it cannot be washed out.   This interaction by interaction proof provides the dirty details for a very simple argument: a system that starts with some kind of parity violation (i.e., different numbers of left and right-handed electrons) cannot have that parity violation removed by QED because QED does not violate parity.  
 
An example of this principle can be found in the experimental set up of a beam of electrons.  Consider an unpolarized beam of electrons travelling along the $z$-direction.  While there are many electromagnetic interactions, the result (\ref{constant}) states that the longitudinal polarization  ${\cal {P}}_z$ remains zero as time passes.  Why does this happen? Consider the collision of two particles in the beam with initial momenta $\vec{k_1}$ and $\vec{k_2}$ and zero total polarization along the $z$-axis.  After the collision polarization can be induced in the transverse direction  ${{\cal \vec{P}}\cdot \vec{k_1}\times\vec{k_2}}$,  but not in the longitudinal directions, ${\cal \vec{P}}\cdot  \vec{k_1}$ or $ {\cal \vec{P}}\cdot  \vec{k_2}$, as this would contradict the fundamental parity symmetry of the QED Lagrangian.  In particular, suppose $ (\vec{k_1}, \vec{k_2})$ are in the $xz$-plane. The polarization can be only induced in the positive $y$-direction.  It is clear that in the thermodynamical equilibrium there will be another pair of particles which produce a polarization in the negative  $y$-direction such that total polarization remains zero. This simple example explains what equation (\ref{constant}) states: if longitudinal polarization of the beam was initially zero, it will remain zero in spite of the fact that each given process may induce polarization.   

So far the discussion has considered only particle-particle QED scattering and not scattering off a background field.  It is tempting to think that the helicity flipping amplitudes of an electron scattering off a semi-classical background field will wash out an asymmetry.  Such arguments neglect that polarization is given to the background field in the form of operators such as the magnetic helicity $\langle \vec{A} \cdot \vec{B}\rangle$ and higher order operators that are not invariant under P transformations.  These operators are the manifestation of the parity lost by electrons to the background field, and is the parity responsible for creating the toroidal magnetic fields discussed in Section 8.  There is a point where the background field, which now contains P-odd parity configurations in the form of magnetic helicity, will start to give parity back to the electrons through scattering.  This steady state is once again achieved through unitarity and thermal equilibrium.  This system of topological currents combined with an E\&M field carrying magnetic helicity is complicated and it is not our goal to present a complete description of how these P-odd effects transform from one form to another.  This would require us to analyse a system comprising of time dependent Maxwell equations with non-static sources. We explicitly calculate the current when these interactions with the external magnetic field are effectively turned off.   However, from the P-odd invariance of QED we know that P-odd effects will stay even when these interactions with the external magnetic field are turned on.   
  
This has a physical analogue in beam physics as well.  As discussed, a beam with no longitudinal polarization that has only particle-particle QED interactions will remain unpolarized.  However, a beam with longitudinal polarization can be created if a magnetic field configuration with nonzero helicity is applied, which is the standard technique used to produce a longitudinally polarized beam.  As this polarized beam propagates it cannot lose its polarization due to the internal particle-particle QED interactions.  If it did, time reversal symmetry, which is unitarity, would tell us that a beam with no longitudinal polarization could spontaneously polarize due to particle-particle interactions. This is untrue and the polarization stays in the beam.  Furthermore, time reversal symmetry applied to the longitudinally polarized beam tells us that when the beam encounters a magnetic field, the field may reconfigure itself to absorb some P-odd magnetic helicity from the beam.  This is exactly how the current in a neutron star imparts magnetic helicity to the field. To conclude, the entire P-odd configuration which includes longitudinal electron polarization and magnetic helicity (and higher order P-odd operators) cannot be washed out by QED interactions.

The final consideration is that some collective bosonic mode such as heat may destroy the current. To destroy the current these dissipative modes must carry the helicity out of the star faster than the current does. It does not imply that P-odd effects are destroyed. Rather it means that P-odd configurations in principle may leave the system. We can account for these dissipative modes by assuming they are carried out of the star by photons or phonons. The helicity modes that the current carries out of the star are subject to direct walk and leave the star close to the speed of light while a photon is subject to a random walk. The number of steps in a random walk is the square of the number of steps in a direct walk, so it takes much longer for a photon to escape than the current to carry helicity out. Some tiny fraction of the helicity manages to escape this way, but the current remains intact. Thermal cooling of the star is also not the dominant cooling mechanism of the star while the current is active, neutrino emission is. The processes that create the neutrinos are the exact processes responsible for the current. We can quantify when the star is too hot for the current to propagate by finding the temperature where $\ell_e < R$. This means that the current can no longer carrying the helicity out of the system and does not flow. Essentially this violates the conditions formulated above for the current to be induced. 

The statement that P-odd  effects, if produced, cannot be destroyed or washed out by conventional QED processes is correct for any system, including condensed matter systems. However, there is a crucial difference between our discussion of the current and conventional condensed matter systems, where it is known that the induced spin polarization inevitably relaxes even though QED preserves parity. In any condensed matter system the relaxation process takes relatively  short period of time as the heat can easily leave the system. This heat is actually represented by  long wavelength photons (microwaves ) that are partly polarized and can easily leave the system. The situation  is drastically  different for neutron stars or quark stars (due to the very large density) when it takes very long period of time before a photon can reach the surface of the star  to have a chance to escape. In principle, the P-odd effects  in neutron/quark  stars will also inevitably relax when no new polarization is pumped in. However, the time scale for this to happen is much longer than in condensed matter systems. This is analogous to the evolution of the magnetic field in neutron stars trapped in a type I superconductor. It is known that the magnetic field should be expelled from the bulk of type-I superconductors. However, it takes a very long time (much longer than the life time of the universe) before the magnetic field is actually expelled from  neutron stars. 

Nowhere in the arguments above have we assumed (or implied) that our state is a pure quantum state with definite parity; on the contrary, our arguments are based on thermodynamics and a density matrix formalism where the polarization for a mixed electron state is defined as the sum over all  particles ${\cal \vec{P}}=\sum_ng_n{\cal \vec{P}}_n $ where $g_n$ is  a probability to find the $n-$th particle with polarization ${\cal \vec{P}}_n $ with normalization $\sum_n g_n=1$.   

For the sake of curiosity there is an example of a strongly interacting, many body system where a P-odd configuration could be produced, but nevertheless is not washed out by very fast strong interactions, which is very similar to the case we consider.  Specifically, we have in mind the charge separation effect in heavy ion collisions \cite{Kharzeev:2007tn,Kharzeev:2007jp} where P and CP odd effect survives (and even have been experimentally observed) in spite of the fact that the system is in thermodynamical  equilibrium with respect to strong interactions. There is a simple argument why a P-odd effect is not washed out by the strong interactions: it is an invariance of QCD with respect to P and CP symmetries. The analogy between our topological current and the P-odd effect in heavy ion collisions is even deeper than it appears:  both effects in fact originate from the same anomaly\cite{Kharzeev:2007tn,Kharzeev:2007jp}.  

\section{Nuclear Matter: The Direct Urca Process}
\begin{figure}[ht]
  \begin{center}
  \includegraphics[width=7cm]{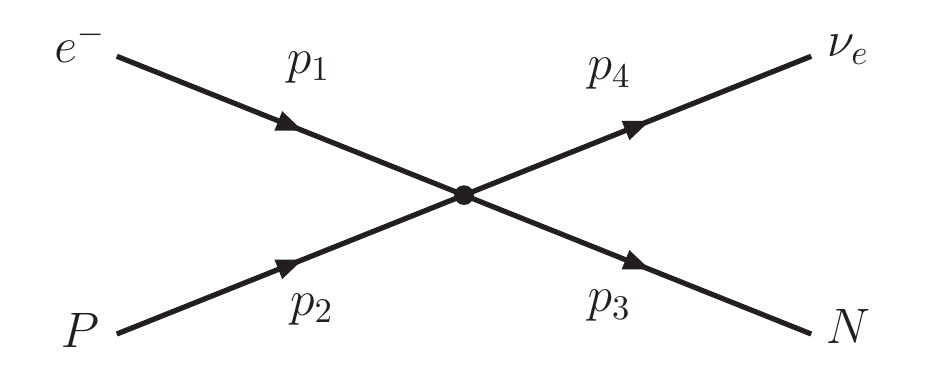}
  \caption{The direct Urca process.}
  \label{fig:feynman_direct}
  \end{center}
\end{figure}

We want to determine the mean free path of the weak interaction of an electron travelling inside a neutron star and the rate at which electrons are created.  To do so we will estimate the transition rate following the standard techniques from \cite{Bahcall:1965zzb}.  Estimating the mean free path allows us   to determine whether the helicity built from the weak interaction is washed out or if the asymmetry can escape the star\footnote{ Here and in what follows we do not assume that an electron physically escapes a star: we use the term  ``escape" to emphasize that the electron can leave the region of degeneracy without being re-scattered by dense surrounding matter. The fate of the  moving electrons when they enter the non-degenerate region $\mu_e\sim T$  from deep degenerate region  depends on specific properties of matter with  $\mu_e\sim T$. In this region the current becomes dissipating, and the electrons  may transfer their energy/momentum to the surrounding dense environment.  This subject  is beyond  the  interests of the present work, and shall not be discussed here. }.   Similar calculations have been done only for neutrinos as the electron's mean free path is assumed to be much shorter due to electromagnetic interactions. However, as we argued above, the  electromagnetic interactions do not wash out P-odd asymmetry and the electrons are allowed to propagate due to the non-dissipating, topological vector current. In order to find the mean free path we will consider the direct Urca process, equation \eqref{eq:Urca}, given by the Feynman diagram in Figure \ref{fig:feynman_direct}. 

Given the order of the diagram this should be to dominant process in normal nuclear matter, but it is not. It is suppressed because the particles taking part are unable to conserve momentum.  If all interacting particles lie on their Fermi surface then,
\begin{eqnarray}
p_e + p_P - p_N \ll T\,,
\end{eqnarray}
where $T$ is  approximately the energy of the neutrino.  In order for the process to conserve momentum the initial electron and proton, or the final neutron, must be far from their Fermi surface in a region with almost no particle occupation.  Forcing this inequality to become an equality introduces a suppression of order $\sim e^{-k_N/T}$. When the proton fraction is sufficiently large, $x_p > 1/9$, this interaction proceeds unfettered \cite{Lattimer:1991ib}.  The transition rate is given by,
\begin{eqnarray}
w = \frac{\Omega^4}{(2\pi)^{12}} \int\! d^3p_1 d^3p_2 d^3p_3 d^3p_4\ S\ (2\pi)^4\,\Omega\ \delta(p_f-p_i)\ |\mathcal{\hat{S}}|^2\,, 
\end{eqnarray}
where $\Omega$ is the volume of the phase space, the $p_i$ are respectively the electron, proton, neutron, and neutrino momentum, $S$ is a statistical factor that takes into account the Fermi blocking, and $\mathcal{\hat{S}}$ the scattering matrix.

 The Fermi blocking factors limit the phase space in which the initial particles can exist and the final particles can be created.  Particles that exist at the beginning of the reaction, such as the electron and proton in this case, are given a factor equal to the Fermi distribution,
\begin{eqnarray}
f_i &=& \frac{1}{1+e^{-(E_i-\mu_i)/T}}\,,
\end{eqnarray}
which tells us where particles exist.  The particles to be created, such as the neutron, are given a factor of one minus the Fermi function,
\begin{eqnarray}
1-f_i &=& \frac{1}{1+e^{(E_i-\mu_i)/T}}\,,
\end{eqnarray}
which restricts the phase space that particles can be created in.  There is no Fermi blocking term associated with the neutrino because they leave the star without interacting and do not become dense enough to form a Fermi surface.   All of these blocking factors make the statistical factor $S$.

We use the standard four Fermi scattering  matrix element,
\begin{eqnarray}
|\mathcal{\hat{S}}|^2 = \frac{G_F^2}{\Omega^4}(1+3C_A^2)\left[1 - \frac{1-c_A^2}{1+3C_A^3}\frac{\bm{p}_1 \cdot \bm{p}_4}{E_1E_4} \right]\,,
\end{eqnarray}
where $C_A = 1.26$ is the Gamow-Teller coupling and $G_F = 1.17 \times 10^{-11}$ MeV$^{-2}$. 

Following \cite{Bahcall:1965zzb} we separate the angular and radial integrals such that the transition rate becomes, 
\begin{eqnarray}
w = \frac{\Omega^5}{(2\pi)^8}|\mathcal{\hat{S}}|^2\,P \,Q\,,
\end{eqnarray}
where,
\begin{eqnarray}
P &=& \int\! p_1^2 dp_1\, p_2^2 dp_2\, p_3^2 dp_3\, p_4^2 dp_4\, S\, \delta(E_f - E_i)\,, \\ 
Q &=& \int\! d\Omega_1\,d\Omega_2\,d\Omega_3\,d\Omega_4 \,\delta^{(3)}(\bm{p}_f - \bm{p}_i)\,.
\end{eqnarray}
We start with the $Q$ integral.  The contribution from the momentum of the neutrino is small compared to the rest so we neglect it in the $\delta$-function.  This allows us to take an integral over all angles causing the term with angular dependence to vanish.  As in \cite{Bahcall:1965zzb} the angular integrals become,
\begin{eqnarray}
\label{eq:Q_direct}
Q= \frac{(4\pi)^3}{2p_1 p_2 p_3}\,.
\end{eqnarray}
We can now do the $PQ$ integral,
\begin{eqnarray}
PQ= \frac{(4\pi)^3}{2} \int\! p_1 dp_1\, p_2 dp_2\, p_3 dp_3\, p_4^2 dp_4\, S\, \delta(E_f - E_i)\,. 
\end{eqnarray}
The first step is to change variables from energy to momentum. In doing so we will approximate the energy by the value at the Fermi surface as these are the particles most likely to participate in the interaction.  For the proton and neutron, which are nearly non-relativistic, we will approximate the energy by the effective mass, $m_N^*\sim 0.8 m_N$ and $m_P^*\sim 0.8 m_N$ \cite{Bahcall:1965zza}.  For the electrons, which are highly relativistic, the chemical potential is equal to the Fermi momentum, $k_e=\mu_e$. These factors can be pulled out of the integral.  Doing the transformation yields, 
\begin{eqnarray}
p_1dp_1 &=& \mu_e dE_1 \\
p_2dp_1 &=& m_P^* dE_2 \\
p_3dp_1 &=& m_N^* dE_3 \\
p_4^2dp_1 &=& E_4^2 dE_4 \,. 
\end{eqnarray}
Performing the neutrino integral, $dE_4$, over the $\delta$-function leaves,
\begin{eqnarray}
\label{eq:PQ_direct}
PQ = \frac{(4\pi)^3}{2} \mu_e m_N^* m_P^* \int\! dE_1\, dE_2\, dE_3\, (E_1+E_2-E_3)^2\, S\,. 
\end{eqnarray}
The next step is to make the integral dimensionless making the substitutions,
\begin{eqnarray}
x_1 &=& (E_1-\mu_e)/T \\
x_2 &=& (E_2-\mu_P)/T \\
x_3 &=& -(E_3-\mu_N)/T\,, 
\end{eqnarray}
such that the statistical factor accounting for the Fermi blocking becomes,
\begin{eqnarray}
S=\prod_{i=1,2,3} \frac{1}{1+e^x}\,.
\end{eqnarray}
The Jacobian of these transformations introduces a factor of $T$ for each measure.  Also, a factor of $T^2$ comes from the $(E_1+E_2-E_3)^2$ term.  The chemical potentials introduced all cancel because of the equilibrium condition $\mu_e + \mu_P -  \mu_N = 0$.  The substitution also causes $m_N^* \rightarrow -m_N^*$ such that we are left with a positive transition rate,   
\begin{eqnarray}
PQ = \frac{(4\pi)^3}{2}\,m_N^*\,m_P^*\,\mu_e\, T^5\, I\,
\end{eqnarray}
where $I$ is an analytic integral,
\begin{eqnarray}
\label{eq:I_3}
I &=& \int_{-\infty}^{\infty}\!dx_1\int_{-\infty}^{\infty}\!dx_2\int_{-(x_1+x_2)}^{\infty}\!dx_3\,\frac{(x_1+x_2+x_3)^2}{(1+e^{x_1})(1+e^{x_2})(1+e^{x_3})} \\
&=& \frac{3}{4} \left(\pi ^2 \zeta (3)+15 \zeta (5)\right)\\
&\approx& 20.56\,.
\end{eqnarray}
Putting everything together, and using the numerical estimates from  the nuclear matter discussion, the transition rate becomes, 
\begin{eqnarray}
\label{eq:rate_direct}
\frac{w}{\Omega} &=&  \frac{G_F^2}{8\pi^5}(1+3C_A^2)\,m_N^*\,m_p^*\,\mu_e\, T^5\, I \\
&=&  3.5 \times 10^{32}\left(\frac{\mu_e}{100 \textrm{ MeV}} \right) \left(\frac{m_N^*}{m_N} \right) \left(\frac{m_P^*}{m_P} \right) \,\left(\frac{n}{n_0}\right)^{2/3}\,(T_9)^5 \textrm{ s}^{-1} \textrm{ cm}^{-3}\,,
\end{eqnarray}
where $T_9 = T/(10^9 \textrm{ K})$ is the dimensionless, scaled temperature.  A typical value for the reduced mass factor is $0.8$. The temperature dependence $T^5$ is consistent with literature---remember that we are calculating the transition rate, not the luminosity.  The luminosity contains an additional factor of energy in the integral that contributes an extra factor of $T$.  

\subsection{Estimate of the current from direct Urca}
The first step is to ensure that the electron can actually escape the star before it decays via the weak interaction.  This involves calculating the mean free path and seeing if it is larger than the radius of the star.  In the literature for calculations of this type, such as neutrino luminosity for cooling, $\Omega$ is the volume of the neutron star---it is necessary to account for all the transitions that occur in the entire star.  Here, we are interested in the decay rate of a single electron, so we will take $\Omega$ to be the volume in which a single electron exists, which is the inverse of the electron number density,
\begin{eqnarray}
\Omega_e = \frac{1}{n_e} = \frac{3\pi^2}{\mu_e^3} =  2\cdot10^{-37} \left(\frac{n}{n_0}\right)^{-2}  \textrm{ cm}^3\,.
\end{eqnarray}

Assuming that the electron travels at the speed of light though the protons---due to the non-dissipative nature of the current with respect to the electromagnetic interactions, the mean free path can be found using,
\begin{eqnarray}
\label{eq:l_direct}
\ell_e \sim \frac{c}{w}\sim 3 \times 10^{9}\ (T_9)^{-5}  \left(\frac{n}{n_0}\right)^{4/3} \textrm{ km}  
\end{eqnarray}
The typical radius for the neutron star is $R\sim 10$ km.  We see that for $T\leq 10^{10} $ K the electrons can easily escape the degenerate region before the P-odd asymmetry  gets washed out due to the weak interactions.  The counterintuitive density dependence---at higher densities the mean free path is larger---is a natural consequence of Pauli blocking.  As the density of the star increases the number of protons increases and we would expect a shorter mean free path, but the number of neutrons increases as well.  The suppression due to the higher neutron chemical potential is greater than the enhancement gained by increasing the proton density.  

Using \eqref{eq:current_estimate} we are now in a position to estimate the magnitude of the current travelling along a single quantum of flux,  
\begin{eqnarray}
\label{eq:current_direct}
\langle j \rangle = 3.0 \times 10^{-8} \, \left(\frac{n}{n_0} \right)^{-2/3}\, (T_9)^5 \textrm{ MeV}\,.
\end{eqnarray}
There is no dependence on the magnetic field because we have normalized per unit of quantum flux.  As discussed earlier this reaction is dominant when hyperons appear at $n/n_0 \ge 3$ or when the proton fraction is large $x_p>1/9$.

From (\ref{eq:l_direct}) we see that is it much easier for an electron to escape a cool star rather than a hot, newly born star, $T\geq10^{11}$ K. If the star is very hot the electron is not able to keep its asymmetry  due to the weak rescattering, as one can see from (\ref{eq:l_direct}).  As it cools there will be a critical value where the electrons can escape, but are still created at a large rate, meaning the current is very large (\ref{eq:current_direct}).  The current is largest when the star is hot, but not so hot that the electrons are unable to escape the region of degeneracy, $\mu_e\gg T$. This temperature is roughly determined by the condition  that the electron mean free path with respect to weak interactions   is approximately equal to radius for the neutron star is $R\sim 10$ km.  It is expected that this temperature  drastically depends on the equation of state and other specific  properties of the environment as  (\ref{eq:l_direct}) suggests.

\subsection{The effect of a large magnetic field on  the transition rate}
In this short subsection we will argue that the effects of the magnetic field can be safely neglected in calculating the transition rate. Of course, the magnetic field still plays a crucial role in producing the required asymmetry for the current, see Section 3 and the Appendix for details. There has been much consideration of Landau levels on the rates of processes that occur in neutron stars.  Landau levels can have drastic effects on the spin   of electrons, though these effects are suppressed by the unusually high chemical potentials found in neutron stars.  An extremely large magnetic field is needed for these effects to manifest themselves in neutron stars.  It must be roughly comparable to the chemical potential $eB\sim \mu_e^2$ in order for substantial changes for the transition rates to occur. Numerically, this corresponds to very large fields $B\sim B_{\textrm{c}}{\mu_e^2}/{m_e^2}\sim 10^{17}$G, which are much larger than the magnetic fields found in typical neutron stars. 

To introduce Landau levels to the earlier calculation we can simply replace the electron dispersion relation with,
\begin{eqnarray}
E_e^2 = p_z^2 + m_e^2(1+2nb)\,,
\end{eqnarray}
where $n$ are all natural numbers, but only electrons with a spin antiparallel to the magnetic field are allowed in the lowest, $n=0$, level and $b=B/B_{\textrm{c}}$ is the magnetic field normalized to the critical field, $B_{\textrm{c}} = m_e^2/e = 4.4 \times 10^{13}$ G.  The the electron phase space becomes,
\begin{eqnarray}
\Omega \int \frac{dp_z}{2\pi} \sum_{n=0}^{n_{\textrm{max}}} \frac{g_n m_e^2b}{(2\pi)^2}\,, 
\end{eqnarray}
where the term after the sum is the degeneracy per unit area and $g_n$, equal to $1$ for $n=0$ and $2$ otherwise, counts the spin degeneracy.  The maximum Landau level occurs for $p_z=0$ when all the energy goes to putting the electron in the Landau level and none goes into the momentum.  This yields,
\begin{eqnarray}
n_{\textrm{max}} = \frac{E_e^2 - m_e^2 }{2m_e^2b}\,. 
\end{eqnarray}
If the number of  Landau levels is very large, $n_{\textrm{max}}\sim \frac{E_e^2}{2bm_e^2}\gg 1$, which is a common case for a typical neutron star, then the phase space returns to the the one we used in the previous section  \eqref{eq:rate_direct}. Therefore, for a typical neutron star   
 the transition rate basically remains the same as the electron phase space essentially unchanged, 
 \begin{eqnarray}
\Omega \int \frac{dp_z}{(2\pi)^3}\mu_e^2 \,, 
\end{eqnarray}
where we used $E_e \sim \mu_e$ and $\mu_e\gg m_e$.  This result is in accordance with our previous  rough argument that a very large magnetic field is required to produce any substantial changes.

We want to contrast this generic case with a rare situation when    only the lowest Landau level is accessible, $n_{\textrm{max}} = 0$.  This occurs when the neutron star is abnormally cold, the magnetic field is unusually large, or the electron chemical potential is unnaturally small \cite{Sagert:2007as}.  In this approximation we recover the usual electron dispersion relation, while  the available phase volume becomes,
\begin{eqnarray}
\Omega \int \frac{dp_z}{(2\pi)^3}m_e^2b\,.
\end{eqnarray}
As before, we move the momentum dependent parts of the phase space outside of the integral by approximating them by their Fermi momenta.  The rest of the integral takes place identically.  We can compensate for the magnetic field in our calculation of the transition rate simply by taking,
\begin{eqnarray}
\label{b}
\mu_e \rightarrow \frac{m_e^2\,b}{\mu_e}\,.
\end{eqnarray}
In fact this transformation can be done to account for the magnetic field for any of the other rates we derive \eqref{eq:rate_modified}, \eqref{eq:rate_kaon}, \eqref{eq:rate_quark}.  In particular, for the direct Urca case we are left with,
\begin{eqnarray}
\frac{w}{\Omega} = \frac{G_F^2}{8\pi^5}(1+3C_A^2)\,m_N^*\,m_P^*\,\frac{m_e^2\,b}{\mu_e}\, T^5\, I,
\end{eqnarray}
which we can compare to equation \eqref{eq:rate_direct}.  This expression is valid in three cases: $T\ll10^9$ K, $\mu_e \ll 100$ MeV, or $b\gg1$.  Only the last of these has a real chance of manifesting itself; magnetars have huge magnetic fields $B\sim 10^{16}$ G.  Otherwise the Landau levels do not have a significant contribution and the earlier expression is valid.

\section{Nuclear Matter: The Modified Urca Process}
\begin{figure}[!ht]
  \begin{center}
  \includegraphics[width=8cm]{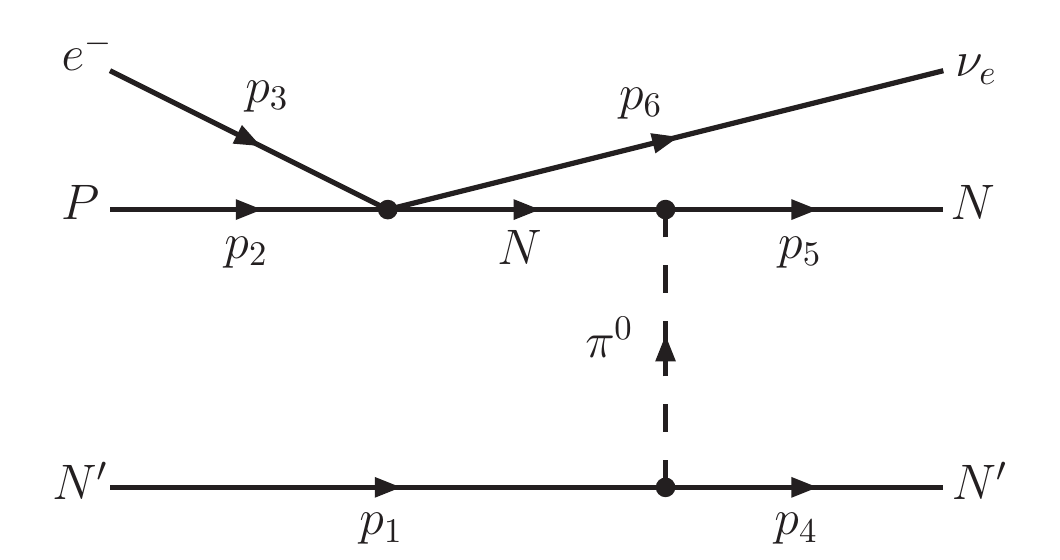}
  \caption{The modified Urca process.}
  \label{fig:feynman_modified}
  \end{center}
\end{figure}

The direct Urca process is the simplest to consider, but it is unlikely to be the most common process in a star.  It is much easier to conserve momentum if a nucleon is included to supplement momentum transfer. We will consider the mean free path of an electron scattering of a proton assisted by a neutron, equation \eqref{eq:Urca}.  The inclusion of the nucleon-nucleon interaction into the matrix element is non-trivial and we will use the one pion exchange/landau liquid method found in \cite{Friman:1978zq}, the interaction illustrated in Figure \ref{fig:feynman_modified}.

There are many other diagrams similar to the one illustrated.  They include all possibilities of pion exchange between protons and nucleons as well as crossing diagrams to account for the final neutrons being indistinguishable. Summing them all causes the vector contributions to cancel and we are left with the scattering matrix approximated to be,
\begin{eqnarray}
\sum_{spins}|\mathcal{\hat{S}}|^2 = \frac{1}{\Omega^6}\frac{32\,G_F^2}{(\mu_e)^2}C_A^2\left(\frac{f}{m_\pi}\right)^4\alpha_\textrm{Urca}\,,
\end{eqnarray}
where $f\sim1$ is the p-wave $\pi N$ coupling constant, $C_A = 1.26$ is the Gamow-Teller coupling, and $\alpha_\textrm{Urca}\sim (0.63-1.76) (n_0/n)^{2/3}$ is a factor that accounts for the pion propagator and the short range Landau liquid contributions.  Following \cite{Friman:1978zq}, but somewhat preempting the calculation, we approximate the propagator of the internal nucleon using the total lepton energy $\mu_e$---remember that the energy the neutrino carries away is negligible. 
 
To calculate the mean free path we will first calculate the transition rate, 
\begin{eqnarray}
w = \frac{\Omega^6}{(2\pi)^{18}} \prod_{i=1}^6 \int \! d^3p_i\ S\ (2\pi)^4\Omega\ \delta^{(4)}(p_f-p_i)\ \sum_{spins}|\mathcal{\hat{S}}|^2\,,
\end{eqnarray}
where $S$ is the Pauli blocking factor.

We separate the transition rate into angular and radial parts leaving,
\begin{eqnarray}
w= \frac{\Omega}{(2\pi)^{14}}\sum_{spins}|\mathcal{\hat{S}}|^2\ P\ Q\,, 
\end{eqnarray}
where,
\begin{eqnarray}
P&=&\prod_{i=1}^6\int\! p^2_idp_i\ S\ \delta(E_f-E_i)\\
Q&=&\prod_{i=1}^6\int\! d\Omega_i\ \delta^{(3)}(\bm{p}_f - \bm{p}_i).
\end{eqnarray}
We start with the $Q$ integral.  The contribution from the momentum of the neutrino is negligible compared to the rest so we neglect it in the $\delta$-function.  This allows us to take an integral over all angles causing the term with angular dependence to vanish.  We then follow \cite{Bahcall:1965zzb} in doing the angular integrals to get,  
\begin{eqnarray}
Q= \frac{(4\pi)^5}{2p_1 p_4 p_5}\,.
\end{eqnarray}
We can now do the $PQ$ integral. Changing variables from momentum to energy and approximating the energy by the particles fermi energy changes the measures to,
\begin{eqnarray}
p_1dp_1&=&m_N^*dE_1 \\
p_2^2dp_2&=&\mu_e\, m_P^*dE_2 \\
p_3^2dp_3&=&\mu_e^2 dE_3 \\
p_4dp_4&=&m_N^*dE_4 \\
p_5^2dp_5&=&m_N^*dE_5 \\
p_6^2dp_6&=&E_6^2dE_6 \,,
\end{eqnarray}
where we took $k_P\sim\mu_e$.  Then we take the $E_6$ integral over the delta function and make substitutions $x_i = \pm\beta(E_i-\mu_i)$ to make the remaining integral dimensionless as in the direct Urca case.  Compared to direct Urca, the two extra particles contribute two more Fermi distributions to the statistical factor, 
\begin{eqnarray}
S=\prod_{i=1}^5 \frac{1}{1+e^x_i}\,.
\end{eqnarray}
Using the beta equilibrium condition, $\mu_P + \mu_e = \mu_N$, all chemical potentials cancel, leaving,
\begin{eqnarray}
PQ= \frac{(4\pi)^5}{2}\ (m_P^*)^2\ (m_N^*)^4\ \mu_e\ T^7\ I\,
\end{eqnarray}
where $I$ is the integral,
\begin{eqnarray}
I&=&\prod_{i=1}^5\left(\int_{-\infty}^\infty\! \frac{dx_i}{1+e^{x_i}}\right) \left(\sum_{j=1}^5x_j\right)^2 \Theta{(x_1+x_2+x_3+x_4+x_5)}  \\
&=&\prod_{i=1}^4\left(\int_{-\infty}^\infty\! dx_i\right) \int_{-(x_1+x_2+x_3+x_4)}^\infty\!dx_5\ \left(\sum_{j=1}^5x_j\right)^2 \prod_{k=1}^5 \frac{1}{1+e^{x_k}} \\
&=& \frac{945}{32}\zeta(7) + \frac{75}{8}\pi^2 \zeta(5)  + \frac{9}{16} \pi^4 \zeta(3) \\
&\approx& 192\,.
\end{eqnarray}
Gathering all the terms together we get the transition rate per unit volume for an electron scattering off a proton assisted by a neutron,
\begin{eqnarray}
\label{eq:rate_modified}
\frac{w}{\Omega} &=& \frac{G_F^2\,C_A^2}{\pi^9}\ \frac{(m_N^*)^3m_P^*\,\mu_e}{m_\pi^4}\ \alpha_\textrm{Urca}\ T^7\ I \\
&=& 9.2 \times 10^{26}\ \left(\frac{\mu_e}{100 \textrm{ MeV}}\right)\,\left(\frac{m_N^*}{m_N}\right)^4\, T_9^7\ s^{-1}\ cm^{-3}\,,
\end{eqnarray}
where as discussed earlier $m^*_N \sim m^*_P \sim 0.8 m_N$.  The rate of this process is much smaller than we found earlier in the direct case \eqref{eq:rate_direct} and the temperature dependence is now $T^7$ rather than $T^5$.  Also, because of the density dependence of $\alpha_\textrm{Urca}$, the transition rate is independent of density.  Both of these differences result from this being a higher order calculation, which involves two extra particles---the ingoing and outgoing neutron assisting in momentum conservation; the transition rate is suppressed by a factor of the QED coupling constant, and the two extra factors of $T$ come from the two extra measures of integration.

\subsection{Estimate of the current from modified beta decay}
We are interested in the mean free path of an electron.  The volume occupied by a single electron is $\Omega_e\sim 10^{-36}$ cm$^3$, and the electron moves at the speed of light, leaving the mean free path to be, 
\begin{eqnarray}
\label{eq:l_modified}
\ell_e \sim 1.4 \times 10^{15}\ (T_9)^{-7} \left(\frac{n}{n_0}\right)^2\textrm{ km.}  
\end{eqnarray}
At the beginning of the star's life, when it is very hot $T\sim10^{12}$ K, the electrons are trapped, then as the star cools the electrons are allowed to escape.  Here we get a similar counterintuitive density dependance as (\ref{eq:l_direct}) but the effect is stronger because there are two neutron Pauli blocking terms to contend with.

We now follow the same prescription as earlier to derive the current,
\begin{eqnarray}
\label{eq:current_modified}
\langle j\rangle = 7.7 \times 10^{-14}\left(\frac{n}{n_0} \right)^{-4/3} \ T_9^{7} \textrm{ MeV.}
\end{eqnarray}
This is by far the weakest current from any phase of matter.  As before the the current is strongest early in the star's life when it is hot, and though it is suppressed it can get quite large at high temperatures due to the $T^7$ dependence. 

\section{Kaon Condensate}
\begin{figure}[ht]  
\begin{center}
  \includegraphics[width=8cm]{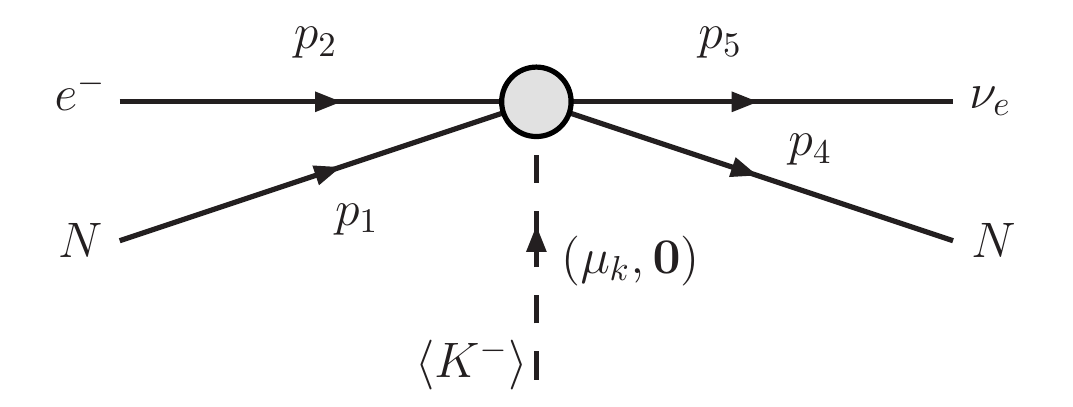}
  \caption{Electron decay in a kaon condensate.}  
  \label{fig:feynman_kaon}
  \end{center}
\end{figure}

As the density of the star gets above three times nuclear density is it possible that a charged kaon condensate will appear \cite{Kaplan:1986yq}.  It is now energetically favourable for electrons to scatter off the condensate and turn into neutrinos.  We are interested in calculating the transition rate of an electron decaying in the presence of a kaon condensate, equation \eqref{eq:kaon}.  The effect of condensates on the scattering matrix was first used to describe pion condensates \cite{Maxwell:1977zz} and later for kaon condensates \cite{Brown:1988ik,Brown:1992ib} and involves evaluating the process given by the Feynman diagram in Figure \ref{fig:feynman_kaon}.

The process needs the aid of a neutron quasiparticle, similar to that of the modified beta decay, but not because it is required to conserve momentum.  We start with the usual hadron and lepton currents and the kaon condensate is described by a chiral rotation in V-spin space, $U=e^{i\gamma_5 \theta}$. For kaons a small rotation can have a large effect so only $\theta\ll 1$ must be considered. The matrix element, as found in \cite{Brown:1988ik}, is given by, 
\begin{eqnarray}
\sum_{\textrm{spins}}|\mathcal{\hat{S}}|^2 = \frac{G_F^2 \theta^2}{4\Omega^4}(1+3C_A^2)\sin^2{\theta_c}\,, 
\end{eqnarray}
where $\theta^2 \sim 0.1$ is the kaon amplitude, $C_A=1.26$ is the Gamow-Teller coupling, and $\theta_c \sim 13^\circ$ is the Cabibbo angle. 

We are interested in the regime where the kaon momentum $\bm{p}_3$ is zero.  The transition rate is given by,
\begin{eqnarray}
w = \frac{\Omega^4}{(2\pi)^{12}} \int\! d^3p_1 d^3p_2 d^3p_4 d^3p_5 S (2\pi)^4\Omega\delta(p_f-p_i)|\mathcal{\hat{S}}|^2\,. 
\end{eqnarray}
where $S$ is the statistical factor containing Fermi blocking terms, $\Omega$ is the volume of the neutron star.  The subscripts on $p_1$, $p_2$, $p_4$, and $p_5$, label the momentum of the ingoing neutron, the electron, the outgoing neutron, and the neutrino, respectively.  Following the work done in \cite{Bahcall:1965zzb} we can separate the angular and radial integrals so each can be evaluated independently,
\begin{eqnarray}
w = \frac{\Omega^5}{(2\pi)^8}\sum_{\textrm{spins}}|\mathcal{S}|^2\,P \,Q\,,
\end{eqnarray}
where,
\begin{eqnarray}
P &=& \int\! p_1^2 dp_1\, p_2^2 dp_2\, p_4^2 dp_4\, p_5^2 dp_5\, S\, \delta(E_f - E_i)\,, \\ 
Q &=& \int\! d\Omega_1\,d\Omega_2\,d\Omega_4\,d\Omega_5 \,\delta^{(3)}(\bm{p}_f - \bm{p}_i)\,.
\end{eqnarray}
We start by doing the angular integral $Q$ is done the same as in earlier cases,
\begin{eqnarray}
Q = \frac{(4\pi)^3}{2p_1p_2p_4}\,.
\end{eqnarray}
The $PQ$ integral can now be started,
\begin{eqnarray}
PQ = \frac{(4\pi)^3}{2p_1p_2p_4} \int\! p_1^2 dp_1\, p_2^2 dp_2\, p_4^2 dp_4\, p_5dp_5\, S\, \delta(E_1 + E_2 - E_3 - E_4 -E_5)\,.
\end{eqnarray}
We first change the variables of integration from momentum to energy $p_idp_i = E_idE_i$ and perform the neutrino integral over the delta function.  In an effort to make the final integral tractable, we follow \cite{Bahcall:1965zzb} in approximating factors next to the measures as constant.  The neutrons are non-relativistic so their energy is just their effect mass, $m^*_N$, and the electron is ultra relativistic, because of its large Fermi momentum, and the energy is just its chemical potential, $\mu_e$.  As before, these factors can then be moved outside the integral.  It is also convenient to change variables to facilitate the final integral over the Pauli blocking factors.  Changing variables takes $(E_1+E_2-E_3-E_4)^2$ to $(x_1 + x_2 + x_4)^2T^2$, where we used the equilibrium condition $\mu_k = \mu_e$.  Upon changing variables the measure of integration gives us a factor $T^3$ and we are left with,
\begin{eqnarray}
PQ = \frac{(4\pi)^3}{2}\,(m_N^*)^2\,\mu_e\, T^5\, I\,
\end{eqnarray}
where $I$ is the same analytic integral from the direct Urca process \eqref{eq:I_3}.

Putting everything together we get the transition rate per unit volume of an electron decaying into a kaon, 
\begin{eqnarray}
\label{eq:rate_kaon}
\frac{w}{\Omega} &=& \frac{G_F^2 \theta^2}{(2\pi)^5}(1+3C_A^2)\sin^2{\theta_c}\,(m_N^*)^2\,\mu_e\, T^5\, I \\
&=& 4.4\times 10^{29} \, \left(\frac{\mu_e}{100 \textrm{ MeV}} \right) \left(\frac{m_N^*}{m_N} \right)^2\left(\frac{n}{n_0} \right)^{2/3} \left(T_9 \right)^5\ \textrm{ s}^{-1} \textrm{ cm}^{-3}\,,
\end{eqnarray}
where $T_9 = T/(10^9 \textrm{ K})$ is the scaled temperature, and $m^*_N \sim 0.8m_N$.  The temperature dependance is the same as the direct Urca process \eqref{eq:rate_direct}, but we would expect it to be of higher order than direct Urca because it involves an extra particle, and thus smaller. The temperature dependence is the same because of the way we treat the kaons as a condensate, rather than an extra particle.  The condensate picture is a rotation of the direct process, rather than a whole new particle interaction as in the modified Urca process. 

\subsection{Estimate of the current in a kaon condensate}
Because of the high chemical potential the electrons must be relativistic leaving the mean free path to be,
\begin{eqnarray}
\label{l_kaon}
\ell_e = 3.0 \times 10^{12} \,(T_9)^{-5} \left(\frac{n}{n_0} \right)^{4/3} \textrm{ km}\,.
\end{eqnarray}
Once again the mean free path larger than the radius of the neutron star and the electrons can escape.    We also notice the counterintuitive, but now familiar, density dependence discussed in the direct Urca case.  This effective mean free path with the helicity intrinsic in the weak interaction creates a current given by equation (\ref{eq:current_estimate}),  
\begin{eqnarray}
\label{eq:current_kaon}
\langle j\rangle \simeq 3.6 \times 10^{-11}\, (T_9)^{5} \left(\frac{n}{n_0} \right)^{-2/3}   \textrm{ MeV.}
\end{eqnarray}
Though the generation of the current in a kaon condensate happens though a wildly different process than direct Urca, and the numbers we use are quite different, we see that after the star has cooled the numbers conspire to give currents of similar magnitude \eqref{eq:current_direct}.      

\section{Quark Matter}
\begin{figure}[ht]
  \begin{center}
  \includegraphics[width=7cm]{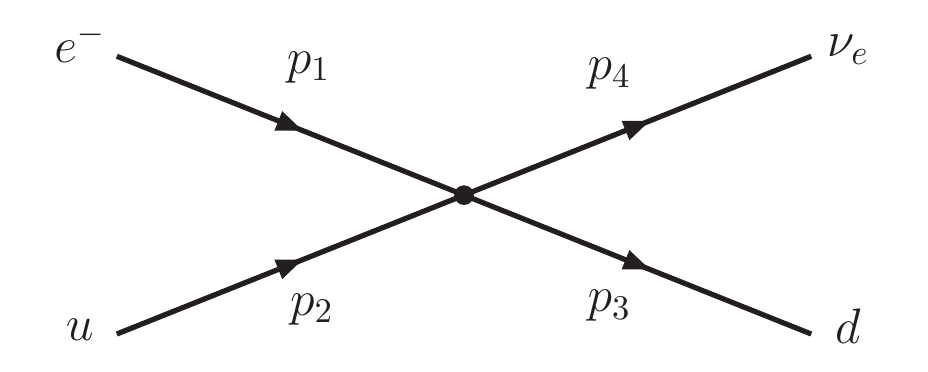}
  \caption{The direct Urca process for quarks.}
  \label{fig:feynman_quark}
  \end{center}
\end{figure}

The last case we consider is what would occur if the hadrons separated into their constituent quarks.  This is the case in quarks stars, where degenerate quarks exist deconfined.  In calculating the mean free path of electrons in quark matter there are two possible reactions we must consider, if we restrict ourselves to the lightest quarks, given in equation \eqref{eq:quark}.  Each is given by the Feynman diagram in Figure \ref{fig:feynman_quark}, where the $d$ quark can be substituted out for the $s$ quark.   

Unlike nuclear beta decay, the lowest order quark beta processes do not require help from an external particle to to conserve momentum; they proceed unsuppressed \cite{Iwamoto:1982}.  Because they are deconfined, the Fermi momentum of the quarks are much closer to each other than the Fermi surfaces of the hadrons in neutron stars.   We will first consider the transition into the down quark.  The transition rate is given by,
\begin{eqnarray}
w = \frac{\Omega^4}{(2\pi)^{12}} \int\! d^3p_1\, d^3p_2\, d^3p_3\, d^3p_4\, S\, (2\pi)^4\Omega\,\delta^{(4)}(p_f-p_i)\,|\mathcal{\hat{S}}|^2\,,
\end{eqnarray}
where $S$ is the statistical factor containing Fermi blocking terms, $\Omega$ is the volume of the phase space.  The subscripts on $p_1$, $p_2$, $p_3$, and $p_4$, label the momentum of the ingoing electron, the up quark, the down quark, and the neutrino, respectively.  The matrix element of this process is found in \cite{Iwamoto:1982},
\begin{eqnarray}
|\mathcal{\hat{S}}|^2 = \frac{6}{\Omega^4} G_F^2 \cos^2\theta_c \frac{16 \alpha_s}{3\pi} \,\left[1-\frac{\bm{p}_3\cdot \bm{p}_4}{E_3 E_4}\right]\,, 
\end{eqnarray}
which looks similar to the usual four Fermi matrix element but contains correction due to the ability to relate the quark's momentum to the quark gluon coupling constant by $\alpha_s = \frac{g^2}{4\pi}$.  We have included the factor to account for the degrees of freedom of the quark in here. The transition rate can be split into two integrals,
\begin{eqnarray}
w = 64\frac{\Omega}{(2\pi)^{9}}G_F^2 \cos^2\theta_c\alpha_s\,P\,Q\,,
\end{eqnarray}
where
\begin{eqnarray}
P&=& \prod_{i=1}^4 \int\! p_i^2dp_i\,S\,\delta{E_f-E_i}\\
Q&=& \prod_{i=1}^4 \int\! d\Omega_i \delta^{(3)}(\bm{p}_f - \bm{p}_i )\left[1-\frac{\bm{p}_3\cdot \bm{p}_4}{E_3 E_4}\right]\,.
\end{eqnarray}
These integrals are nearly identical to the kaon case.  The $Q$ integral is the same as in the direct Urca case \eqref{eq:Q_direct}.  We make an approximation to the remaining $PQ$ integral when we change variables from momentum to energy.  As we have done in previous integrals, the the momentum of the electron and quarks is replaced by their value at the fermi surface.  Performing the neutrino integral, $dE_4$, over the delta function leaves us with the same integral as equation \eqref{eq:PQ_direct}.  We perform a similar change of variables as before to make the integral dimensionless.  Following the same procedure as the direct Urca case, the quark beta equilibrium condition $\mu_e+\mu_u=\mu_d$ causes the chemical potentials to cancel, and the $PQ$ integral becomes,
\begin{eqnarray}
PQ = \frac{(4\pi)^3}{2} \,\mu_e\, k_u\,k_d\, T^5 \,I \,,
\end{eqnarray}
where $I$ is the same integral given in equation \eqref{eq:I_3}, which is the same as the direct Urca and kaon cases. An identical calculation can be done for the electron scattering into strange quarks.  Putting everything together we get the transition rate for electrons in quark matter,
\begin{eqnarray}
\label{eq:rate_quark}
w_d &=& \Omega \frac{4}{\pi^6}\,  G_F^2\, \cos^2\theta_c\,\alpha_s \,k_e\, k_u\,k_d\, T^5 \,I\\
w_s &=& \Omega \frac{4}{\pi^6}\,  G_F^2\, \sin^2\theta_c\,\alpha_s \,k_e\, k_u\,k_s\, T^5 \,I\,.
\end{eqnarray}
To estimate the the magnitude of the transition we assume that $k_d \approx k_u \approx k_s$ as in the massless noninteracting case and use the relationship $k_e \approx \mu_e = (3Y_e)^{1/3}k_q$,  where $Y_e = n_e/n_b$ is the ratio of electrons to baryons in the star.  The Fermi momentum of the quarks is estimated using \eqref{eq:k_quark}.  We follow \cite{Iwamoto:1982} in estimating $\alpha_s \approx 0.4$ and $Y_e \approx 0.01$

The total transition rate is the sum of $w_s$ and $w_d$,
\begin{eqnarray}
\frac{w}{\Omega}= 1.8 \times 10^{31}\, \left(\frac{n_b}{n_0}\right)\, (T_9)^5 \textrm{ s}^{-1} \textrm{ cm}^{-3} \,.
\end{eqnarray}
We see the $T^5$ temperature behaviour that has become a signature for the first order, four Fermi interaction.   
\subsection{Estimate of the current in quark matter}
As before, the first step in finding the current is estimating the mean free path of the electon in quark matter by assuming that the electron propagates at the speed of light.  The same physics creates the current in quark matter and it is still non-dissipating.  We now use the new definition of $n_e$ to determine the volume occupied by a single electron, $\Omega = (Y_e\, n_b)^{-1}$, and we are left with,
\begin{eqnarray}
\label{eq:l_quark}
\ell_e = \frac{c}{w} = 2.9 \times 10^{10} \, (T_9)^{-5} \textrm{ km} 
\end{eqnarray}
The radius of the star, $R\sim10$ km, is much smaller than this and the current will propagate.   In quark matter the mean free path is not dependent on density; the Pauli suppression and enhancement cancel each other---see section 4.1 for a brief discussion.  When the electrons reach the crust they are removed from the system creating a new effective mean free path for the electron.  This effective mean free path with the helicity intrinsic in the weak interaction creates a current,          
\begin{eqnarray}
\label{eq:current_quark}
\langle j\rangle = 2.8 \times 10^{-9} \,(T_9)^{5} \left(\frac{n_b}{n_0}\right)^{1/3} \textrm{ MeV}\,.
\end{eqnarray}
The typical density for quark matter is $n_b \sim 10\, n_0$, but could easily be higher.  Once again the numbers have conspired and the magnitude of the current is close to the value for both direct Urca \eqref{eq:current_direct} and kaon \eqref{eq:current_kaon} processes.  They are all first order processes, but we see a critical difference in the density dependence.  Unlike the other currents the quark current actual gets larger with increasing density.  This happens because quark stars remain charge neutral in a fundamentally different way than neutron stars (see Section 2.4) and the electron chemical potential is determined differently. 

\section{Applications in Neutron Stars}

In this section we discuss five different applications of the topological vector current introduced in this paper,
\begin{eqnarray}
&&\textrm{Section 8.1: Neutron star kicks}\nonumber \\
&&\textrm{Section 8.2: Pulsar jets }\nonumber\\
&&\textrm{Section 8.3: Toroidal magnetic field}\nonumber \\
&&\textrm{Section 8.4: Magnetic helicity}\nonumber \\
&&\textrm{Section 8.5: Type-II superconductivity and precession }\nonumber
\end{eqnarray}
These phenomena are observed in many neutron stars and appear to be unrelated to each other, but we will argue that they originate from the same source---non-dissipating topological currents. All these phenomena have one thing in common: they are P-odd effects.  

We have calculated the magnitude of the topological current in four types of nuclear matter and summarize them in Table \ref{table:summary}.   Direct beta decay, which is also the preferred reaction when hyperons are present, kaon condensates, and quark matter,  all are radically different, but they all make a current about the same order. There is a stark difference in the modified Urca process, which occurs in ordinary nuclear matter where the proton fraction is below $1/9$.  The current is four orders of magnitude smaller than the rest.  This is because it is a higher  order process involving more particles.  It  results in a more severe dependance on temperature which suppresses the process when the star cools.   

\begin{table}
\caption{\label{table:summary} A summary of the current per magnetic flux quantum calculated from each interaction.  The value is for a single quantum of flux, such as that found in a type-II vortex.  The appropriate density for the kaon and the direct Urca currents is $n = 3\,n_0$, for quarks is $n_b = 10\,n_0$, and for modified Urca is $n=n_0$. }
\ra{1.3} 
\begin{center}
\begin{tabular}{@{}lll@{}}\toprule 

& Current $\langle j \rangle$ (MeV)& \\ 
\midrule 
      Direct Beta Decay & $3.0 \times 10^{-8}  \, \left(\frac{n}{n_0} \right)^{-2/3} \, (T_9)^5$  \\
      Modified Beta Decay & $7.7 \times 10^{-14}\left(\frac{n}{n_0} \right)^{-4/3}  \, (T_9)^{7} $  \\
      Kaon Condensate & $3.6 \times 10^{-11}\left(\frac{n}{n_0} \right)^{-2/3}  \, (T_9)^{5} $ \\
      Quark & $2.8 \times 10^{-9} \left(\frac{n_b}{n_0}\right)^{1/3} \, (T_9)^{5}  $   \\
\bottomrule 
\end{tabular} 
\end{center}
\end{table} 

The estimates of the current are relatively small in magnitude, but much larger currents exist on microscopic scales. The estimates represent a component of the current that produces a coherent effect in the entire region where matter is degenerate and the magnetic field all points in one direction. As a technical  remark, the estimates of the current $\langle j\rangle$  are presented in MeV units.  The current is  defined as a  number of particles crossing the surface equivalent to one quantum of magnetic flux per unit time.  We can obtain the electromagnetic current in conventional units by multiplying by $e=\sqrt{4\pi\alpha}$ and converting the result into Amperes: $e \langle j\rangle \sim 10^2$ A for $\langle j \rangle=1$ MeV.

\subsection{Neutron star kicks}
We will briefly discuss the application of topological currents to neutrons star kicks.  A much more detailed analysis can be found in \cite{Charbonneau:2009hq}. It is accepted that pulsars have much higher proper velocities than their progenitors, some moving as quickly as $1000$ km/s \cite{Lai:2003hm,Gunn:1970,Lyne:1994az,Hansen:1997zw,Cordes:1997mm,Fryer:1997nu,Arzoumanian:2001dv,Fryer:2003tc}.  Such large velocities are unambiguously confirmed with the model independent measurement of pulsar B1508+55 moving $1083^{+103}_{-90}$ km/s \cite{Chatterjee:2005mj}. There also appears to be a strong correlation between the direction of the kick and the rotation axis of the star \cite{Wang:2006zia}.  If the electrons carried by the current can transfer their momentum into space the current could slowly and steadily pushing the star over time, much like a rocket.  This is contrary to other mechanisms where the kicks happens shortly after the star's birth. The sustained nature of the kicks allows a rotation-kick correlation $\langle  \vec{P}\cdot \vec{\Omega} \rangle$ to form regardless of the angle between the star's rotation and magnetic field.  Neutron stars spin on the order of milliseconds and the kick occurs over hundreds of years.  The force vectors from the kick form a cone aligned with the axis of rotation that average to produce a kick along the axis of rotation.   Long duration kicks are supported by the analysis in \cite{Spruit:1998}.  

Currently no mechanism exists that can reliably kick the star hard enough.  Explosions during collapse can only reliably kick a star $50$ km/s \cite{Lai:2003hm}, asymmetric explosions can only reach $200$ km/s \cite{Fryer:2003tc}, and asymmetric neutrino emission is plagued by the problem that at temperatures high enough to produce the kick the neutrino is trapped inside the star \cite{Sagert:2006yn,Sagert:2007ug}.  More exotic mechanisms require exotic particles and more fine tuning.  We provide a rough calculation demonstrating that topological kicks can provide the momentum required and do so because of the immense electron chemical potential, not because of the temperature of the star.  A detailed calculation of the kick is presented in \cite{Charbonneau:2009hq}. Also, because of the correlation between the magnetic field and the spin axis, topological kicks naturally align with the spin axis of the star.  

Though it is unclear what will happen to the electrons when they reach the surface of the star, we can estimate the size of the kick if we assume the entire  momentum carried by the topological current will be transferred into space by some means. This assumption is likely correct for  bare quark stars where the crust (the region where $\mu\sim T$) is only about 1000 fm wide \cite{Alcock:1986hz}\footnote{ It has been recently argued that a crust in quark stars could be much larger in size than previously thought due to development of a new  heterogeneous mixed phase \cite{Jaikumar:2005ne, Alford:2006bx}. As we mentioned above, it is not our goal to discuss the interaction of current with the crust, however, an intense current from the core of quark star may destroy the crust in this new    mixed phase in few   locations similar to volcanoes on earth.}   and is most likely wrong for typical neutron stars where the crust is about 1 km thick.  The energy of the electrons would be absorbed by the crust and would not contribute to the kick.  As a consequence we conjecture  that stars with very large kicks, $v\gg 200$ km/s, are quark stars and that slow moving stars, $v \leq 200$ km/s, are kicked by some other means  \cite{Fryer:2003tc} and are typical neutron stars.  Confirmation of this classification would provide a simple and well formulated principle that distinguishes quark stars from typical neutron stars and results in a bimodal distribution for kick velocities as supported in \cite{Arzoumanian:2001dv}. As is known, other criteria such as mass, size, cooling rate, etc.\ cannot easily discriminate  between quark stars and neutron stars, see e.g. \cite{Page:2006ud,Reddy:2002ri} for review. 

The first step in estimating the magnitude of a topological kick is determining the total momentum transferred to the star. There are $N_{\textrm{v}} \sim 7\cdot10^{33}\ B/B_{\textrm{c}}$ quantum units of flux in a star that can be distributed in either superconducting domains or vortices, see eq. (\ref{N}).  The current is independent of the internal structure of the star.  If all the electrons are shot out of the star or transfer their entire momentum, then the momentum is given by the total current (\ref{current-new}), the number of electrons that leave the star per unit time.  The momentum a single electron transfers out of the star is equal to its Fermi momentum, $k_e = 100$ MeV. The fuel for the kick is the chemical potential, not the temperature. Therefore the kick may continue even when star is already cool; the kick in our mechanism is not an instant event, but is rather a long, slow, steady process that pushes the star.  Putting this all together, the current transfers $N_{\textrm{v}}\,k_e\,\langle j\rangle$ units of momentum per unit of time.  

The appropriate way to estimate the magnitude of the kick is to integrate the rate of momentum transfer per unit   time by  taking into account the time evolution of the star as the current is very sensitive to the temperature/time as Table 1 demonstrates. The rate of momentum transfer  also changes as the star cools and  the environment  (density, phase) changes. We neglect all these complications and take a value for the current corresponding to $T\sim10^9$ K, which is far from  the maximum current can potentially be induced. 

The momentum transfer required for the star to reach a velocity of $v = 1000$ km/s is quite large.  Per baryon, the momentum required is $m_n\,v \sim 3$ MeV.  The total baryon number of a neutron star of one solar mass is $B_n \sim 10^{57}$.  The momentum for the entire star is then $P \sim 3\cdot 10^{57}$ MeV. If we choose a current $\langle j \rangle = 10^{-10}\, (T_9)^5$ MeV, corresponding to an average current in Table \ref{table:summary}, the time required to attain this momentum is,
\begin{eqnarray}
t = \frac{P}{N_{\textrm{v}}\,k_e\,\langle j\rangle} &\approx& 3\cdot10^{10} \left(\frac{v}{1000 \textrm{ km/s}}\right) \left(\frac{B_{\textrm{c}}}{B}\right)\left(\frac{10^{10}\textrm{ MeV}}{\langle j \rangle}\right)\left(\frac{10^{9} \textrm{ K}}{T}\right)^5 \textrm{s} \nonumber \\
&\approx& 10,000 \textrm{ years,}
\end{eqnarray}
We restore the canonical dimensionality  of MeV$^{-1}$ in seconds by multiplying $\hbar=6.58 \cdot 10^{-22}$ MeV$\cdot$s. This is a conservative estimate of the kick strength.  The current is much larger at the star's birth (when it is hot) and kicks such as those seen in, for example, the Vela pulsar can be easily explained.  If electrons actually leave the star,  rather than transferring their momentum through radiation, the electron chemical potential will slowly decrease and the current may stop running.  Charge neutrality will cause matter to accrete isotropically and possibly maintain some of the chemical potential. 

As previously mentioned it is unlikely that the electrons could escape a typical neutron star---the crust has a thickness on the order of $1$ km  and electrons leaving the degenerate core would quickly be absorbed.  On the other hand, quarks stars have a thin interface between the degenerate matter and open space \cite{Alcock:1986hz} that electrons might be able to easily penetrate, or transfer their momentum by emitting photons into space. This difference in crusts could provide a valuable distinction: stars with large kicks are quark stars, stars with small kicks are neutron stars. 

Our kick mechanism is similar to neutrino driven kicks in that both use P-odd effects and particles to carry momentum out of the star.  The electron and neutrino have similar mean free paths, but the electrons leave much more slowly than the neutrinos.  The fundamental difference in the two carriers.  At low temperatures, when the neutrinos can escape the star, they do not carry enough energy to explain the kick.  In contrast, the electrons that make the non-dissipating current (or more precisely the quasi-particles which freely travel along the magnetic field) carry very large momentum $\sim \mu_e$.  As a result the neutrino carries too little momentum when the mean free path becomes sufficiently long, while the momentum carried by the topological current remains very high even at very low temperatures $T\sim 10^8$ K. This is the advantage of our kick mechanism.

\subsection{Pulsar jets}
A different but likely related phenomena is the recent observation of pulsar jets \cite{1538-4357-554-2-L189} that are apparently related to neutron star kicks \cite{Wang:2005jg, Lai:2000pk}.  It has been argued that spin axes and proper motion directions of the Crab and Vela pulsars are aligned.  Such a correlation would follow naturally if we suppose that the kick is caused by a non-dissipating current as we mentioned above.   The current, and thus the proper motion, is aligned with the magnetic field, which itself is correlated with the axis of rotation.  It would be very tempting to identify the observed inner jets \cite{1538-4357-554-2-L189} with the electrons/photons emitted as a result of the induced current.     In this sense the mechanism for the kick is similar to the electromagnetic rocket effect suggested previously \cite{Lai:2000pk}.

It is possible that evidence of the topological current may be directly detected in these jets. An observational consequence of the current is that a component of the X-ray emission in the trail of the neutron star will be left circularly polarized.  Because of the parity violation in the star, only left-handed electrons will contribute to the kick.  When these left-handed electrons interact they will create mostly left-handed photons, which coherently will be seen as left circularly polarized X-rays.  As there are many other sources of X-rays the contribution is likely very small, but it further motivates the need for higher precision X-ray polarimetry \cite{Weisskopf:2006xb}.         

\subsection{Toroidal magnetic fields}
There is a strong theoretical evidence for the existence of toroidal fields in neutron stars  based on the stability of the poloidal magnetic field.  References \cite{Markey:1973,Wright:1973,Flowers:1977,Prendergast:1956,Tayler:1973} argue that toroidal and poloidal fields of similar magnitudes must be necessary to stave off hydrodynamic instabilities---the toroidal field suppresses poloidal instabilities and vice versa.  Much has been done to find the observational consequences of a toroidal field, for example \cite{Page:2007br}.   

Estimating the toroidal magnetic field is a very complicated problem that requires
a self consistent solution of the equation of the magnetic hydrodynamics.  Our induced, topological currents represent  only a small part of the system. We are not attempting to solve this problem. Instead, we shall argue that the currents we estimated are more than sufficient to induce the toroidal magnetic field correlated on large scale of order 10 km.

A natural consequence of having a current running parallel to the poloidal magnetic field is that a toroidal component $H_{\textrm{tor}}$ will be induced.  The size of the field can be calculated naively  using Ampere's law, but there is a subtlety because the magnetic field is being induced inside a superconductor.  The magnetic field observed in neutron stars $B\sim10^{12}$ G is actually induced by a much larger field $H$.  The suppression comes from the perfect diamagnetism of the proton superconductor  (the Meissner effect).  This perfect diamagnetism is ruined at a critical field $H_{\textrm{c}} \sim (\Phi_0/4\pi\lambda^2)\sim 10^{15}$ G where flux penetrates the star through small regions where superconductivity has been destroyed (vortices or domains).  The supercurrents responsible for the perfect diamagnetism do not flow as easily and a small field is induced.

Regardless whether the flux penetrates the superconductor as single vortices or in flux domains, we can assume that the superconductor is type-II.\footnote{The mechanism for type-I like superconductivity discussed in \cite{Charbonneau:2007db} relies on the electromagnetic interaction between currents carrying vortices, not in altering the value of the Landau-Ginzburg parameter $\kappa=\lambda/\xi$.  We then still use results from type-II superconductors (indicated by $\kappa >1/\sqrt{2}$) but the vortices are now bunched together in large domains with higher winding numbers.}  The relationship between the applied magnetic field $H$ and induced magnetic field $B$ in a type-II superconductor is very nonlinear.  The details have been worked out in \cite{Tinkham:1996} and \cite{PhysRevD.16.275}, where the latter is a direct application to neutron stars.  The important points are that below the first critical field $H<H_{c1} $  there is no magnetic field $B$ induced.  Just above the critical field a magnetic field appears that is approximately $B\sim10^{-3} H_{c1}$.  As the applied magnetic field is increased above $H_{c1}$ the induced field starts to approach the applied field.

We want to determine if the topological current produced by the poloidal field can induce a sufficient toroidal field by finding the length scale where $H_{tor} \sim H$.  Following \cite{PhysRevD.16.275} we assume that $H\sim H_{\textrm{c}}$ and we get the relationship $B\sim 10^{-3}H$ for our magnetic field.  We apply Ampere's law for a region of size $L$, 
\begin{eqnarray}
\label{Ampere}
 H_{\textrm{tor}}2\pi L = e j\cdot\left(\frac{\pi L^2 B}{\Phi_0}\right),
\end{eqnarray}
where the expression in brackets describes the number of unit fluxes bundled in the area $\pi L^2$ such that we get the total current enclosed in our loop.  We take $\Phi_0=\pi/e$ and substitute use our relationship between $H$ and $B$. 
 
The naive estimate leads to the following expression for $H_{\textrm{tor}}$ in terms of   magnitude of poloidal magnetic field $H$, 
\begin{eqnarray}
\label{Ampere-1}
 \frac{H_{tor}}{H} \sim \alpha \langle j \rangle   L \sim  4  \left(\frac{\langle j \rangle}{10^{-10}\text{ MeV}}\right)
 \left(\frac{L}{ \text{km}}\right).
\end{eqnarray}
This shows that a typical current from Table \ref{table:summary} can induce a toroidal field the same magnitude as the poloidal field on scales the order $L\sim 1$ km, within the typical size of a neutron star.  It is quite obvious that  our estimate becomes unreliable when $H_{\textrm{tor}}\geq H$ and we can no longer ignore the current induced by the toroidal field.  For  $H_{\textrm{tor}}\geq H$ the problem  requires a self consistent analysis which is beyond   scope of the present paper. The point is that the toroidal field  obviously develops as a result of topological currents and eq. (\ref{Ampere-1})  shows that its magnitude can easily become the same order  as the poloidal magnetic field. If superconductivity is completely destroyed, as we will discuss Section 8.5, $B=H$ and the toroidal field can induced on a much smaller scale, $L\sim1$ m.

\subsection{Magnetic helicity}
Magnetic helicity, see e.g. \cite{Choudhuri:1998}, is defined as, 
\begin{eqnarray}
\label{mhelicity}
\mathcal{H}\equiv\int \!d^3x \,\vec{A}\cdot\vec{B}.
\end{eqnarray}
The magnetic helicity is a topological object that can be expressed in terms of the linking number $n_{\left( \gamma _{1},\gamma _{2}\right)}$ of  two curves $\gamma_1$ and $\gamma_2$. The precise relation between $\mathcal{H}$ and interlinked flux $\Phi_1$ and $\Phi_2$ is given by,
\begin{eqnarray}
\label{mhelicity1}
\mathcal{H}= 2\Phi_1 \Phi_2=2\Phi_0^2\,  N_1 N_2 
\end{eqnarray}
where $\Phi_1=\Phi_0 N_1$ and $ \Phi_2=\Phi_0 N_2 $ are expressed in terms of unit flux $\Phi_0$ and the linking number is simply reduced to $n_{\left( \gamma _{1},\gamma _{2}\right)}= N_1N_2$. Therefore $\mathcal{H}$ takes integer values up to a normalization $2\Phi_0^2$.  This linking number is  preserved, $\frac{d\mathcal{H}}{dt}=0$, in a magneto-fluid with zero resistivity, which is a very good approximation for neutron stars.  This topological invariance provides the stability necessary for the poloidal field.

We want to emphasize that the magnetic helicity  is the dot product of a vector and a pseudovector, making it a pseudoscalar.  Under the parity transformation $\vec{x}\rightarrow -\vec{x}$ the magnetic helicity is P-odd:  $\mathcal{H}\rightarrow -\mathcal{H}$.  This implies that  the magnetic helicity 
can be only induced if there are parity violating processes producing   a  large coherent effect on macroscopic scales.  Many attempts to generate helicity rely on instabilities in the magnetic field caused by the star's rotation.  Such correlations $\vec{B}\cdot\vec{\Omega}$ are P-even, and though they may generate toroidal fields they cannot be responsible for helicity. 

Our observation here is that the non-dissipating topological current introduced in  the present work
has precisely this property: the topological current produces the P-odd correlation $\langle \vec{P}\cdot \vec{B}\rangle$ and is capable of inducing magnetic helicity $\mathcal{H}\sim j$.  In fact, our estimate for the induced toroidal field $H_{\textrm{tor}}$ unambiguously implies that the magnetic helicity will be also induced, see eq. (\ref{Ampere-1}).  The magnetic flux from the toroidal and poloidal fields is always  interlinked and contributes to the magnetic helicity, 
\begin{eqnarray}
\label{mhelicity2}
\mathcal{H}=2 \Phi_{\textrm{torr}}\Phi, 
\end{eqnarray}
where $\Phi$ and $\Phi_{\textrm{torr}}$ are the original poloidal and induced toroidal magnetic fluxes (\ref{Ampere-1})  correspondingly.  

Strong observational evidence, see \cite{Page:2007br} and references therein, supporting the presence of the toroidal component unambiguously suggests that the magnetic helicity $\mathcal{H}$ must be non-zero in neutron stars.  The P-odd quality of the magnetic helicity may be strong, indirect evidence supporting our claim that P-odd topological currents have been induced at some moment in the star's life.  Otherwise, it is very difficult to understand how such a large, coherent P-odd effect could be produced.         

\subsection{The conflict between vortices and precession}
It has been observed that neutron stars precess \cite{Stairs:2000,shabanova:321} and that the degree of precession conflicts with the commonly held belief that the protons form a type-II superconductor in the core \cite{Link:2003hq,Link:2006nc}.  It has been shown that if the superconductor is type-I there is no conflict \cite{Sedrakian:2004yq}.

When a magnetic field is applied to a type-II superconductor the flux finds it energetically favourable to penetrate it by forming many vortices each carrying a unit of quantum flux.  In a neutron star the large number of these get tangled with the superfluid neutron vortices that have formed to carry angular momentum.  If the star precesses with a large enough angle the superfluid vortices must break through the superconducting vortices for rotation to continue.  Incredibly large amounts of energy are dissipated in this process which would cause the star to stop rotating.  We conclude then that the superconductor must be type-I where the flux bunches in large groups organizing macroscopically large domains  and there is room for the neutron vortices to move around.  

The problem is that the Landau-Ginzburg parameter for a typical neutron star indicates the superconductor is type-II.  A solution to this is suggested in \cite{Charbonneau:2007db}.  If a sufficiently large current runs along type-II vortices an attractive force arises that causes the vortices to bundle together like they would in a type-I superconductor even though the Landau-Ginzburg parameters indicate type-II behaviour. 

Even if the current is not strong enough to make the vortices attract each other it has also been argued \cite{Charbonneau:2007db} that the mere presence of an induced, longitudinal current, arbitrarily small, would destroy superconductivity, thus resolving the problem. In many condensed matter systems such kind of instability has been experimentally tested,  see  \cite{Charbonneau:2007db} for relevant references on test of this instability in condensed matter systems.  This instability can be delayed for small currents or even stabilized due to impurities.  But the lesson from these condensed matter systems is that when a current aligns with the magnetic field creating the vortex the properties of the vortex lattice are completely changed or destroyed.
 
We expect similar behaviour in regions of the neutron star where both the Landau-Ginzburg parameter suggests type-II behaviour and longitudinal currents are induced.  While many features of the system are still to be explored, the point is that that even small topological currents along the magnetic field will likely destroy the vortex lattice by replacing it with a new, unknown structure, similar to the condensed matter experiments mentioned above.  The exact state is not essential at the moment, only that the Abrikosov lattice is destroyed by longitudinal currents and the conflict formulated in \cite{Link:2003hq,Link:2006nc} is resolved.  

\section{Conclusion}
The goal of this paper is to argue that a persistent, topological current is induced in neutron stars.  All the requirements are present: a large degeneracy $\mu_e\gg T$,  an approximately  chiral, Dirac-like  spectrum at $\mu_e\gg m_e$, and a large magnetic field $B$. It is an unusual but beautiful pure, quantum phenomena that has no analogue in classical physics.   It is fundamentally new, which raises a question: have similar phenomena been studied? There are systems with the potential to manifest anomalous currents.  More so, effects similar to those discussed in this paper have been experimentally tested in some condensed matter systems and are going to be tested at RHIC and there is a possibility that these effects can be experimentally tested in terrestrial laboratories. 

In condensed matter laboratories low temperatures and strong magnetic fields present no technical difficulties.  The key is finding a system of quasiparticles with a Dirac-like spectrum.  There are such systems: superfluid $He^3$, high $T_{\textrm{c}}$ superconductors with d-wave pairing,  and graphene.  Remarkably, in superfluid $He^3$ the current analogous to our anomalous current has been observed, see reviews \cite{RevModPhys.59.533} and \cite{Volovik2001195}.

The relativistic heavy ion collider (RHIC) at Brookhaven also has potential.   An analogue to the anomalous current has been used to predict a charge separation effect \cite{Kharzeev:2007tn} and preliminary experimental results are supporting it.  In this analogue each requirement for the current to exist in the neutron star has its complement: the role of the coherent magnetic field is played by the angular momentum $\vec{L}$, which occurs at non-central nuclei collisions, and the the role of parity violating effects is played by the induced $\theta$ vacua.  The observation of the charge separation indirectly supports our prediction of induced anomalous currents.  We should remark here that very strong interactions in each nuclei-nuclei collision event cannot wash out the produced asymmetry. This parallels our argument in Section 3 that strong electromagnetic interactions do not wash out the P-odd produced asymmetry and the relevant scale of the problem is the mean free path of the electron due to the weak interactions that are capable of washing out P-odd effects. 

Confirmation of our claim that anomalous currents exist in neutron stars would have an enormous effect on the physics of neutron stars.  In particular, it may explain neutron star kicks and pulsar jets (see Sections 8.1 and 8.2) and give a way to discriminate between neutron stars and quark stars, it may shed light on the nature of the toroidal magnetic field required for stability of the poloidal field (see Section 8.3), and it may resolve the conflict between the observed precession of a neutron star and type-II superconductivity commonly believed to exist in the core (see Section 8.5).  Topological currents also provide a source of finite magnetic helicity, a P-odd topological invariant that does not decay in a neutron star environment.  This may shed some light on the origin of the strong, self-supporting system of toroidal and poloidal magnetic fields in neutron stars  (see Section 8.4).  This necessity of finite magnetic helicity is strong, indirect evidence that non-dissipating currents move along $\vec{B}$.
 
In Section 8 we mentioned many apparently unrelated observational effects: neutron star kicks, toroidal fields, and magnetic helicity.  These all have a P-odd symmetry and it is likely that they all originate from the same P-odd physics.  The topological vector current introduced in this paper occurs because of parity violating effects.  This current may be responsible for all of these P-odd phenomena.   
 
\acknowledgments
The authors would like to thank Lionel Brits for countless discussions; Jeremy Heyl, Kelsey Hoffman, Jean-Fran\c{c}ois Caron, and Saul Davis for their feedback on the manuscript; Dam Thanh Son,  Misha Stephanov, Cliff Burgess, and Bill Unruh for useful discussions and comments.  This work was supported in part by the National Science and Engineering Council of Canada.

\appendix
\section{Calculation of the Helicity $\Lambda(\mu,B,T)$ }
In order to calculate the average helicity of the electrons, we must consider the creation of an electron through nuclear beta decay in a large magnetic field and chemical potential.  There have been many calculations of transition rates in magnetic fields, notably \cite{Matese:1969zz,Dorofeev:1985az}, and more recently for both large magnetic fields and chemical potentials \cite{Sagert:2007as} and references within.  We are looking for a specific property of the weak interaction, the helicity of electrons produced in a large magnetic field and chemical potential.  The helicity of a particle is given by $\Lambda=\bm{\sigma}\cdot \mathbf{p}/|\mathbf{p}|$, where $\bm{\sigma}$ are the Pauli matrices and $\mathbf{p}$ is the particle's spatial momentum.  The two eigenstates of the helicity operator correspond to the values 
\begin{equation}
\mathbf{p}\cdot \boldsymbol{\xi} = \pm |\mathbf{p}|\,,
\end{equation}
where $\boldsymbol{\xi}$ is the rest spin vector of the particle.  The expectation value of the helicity can be calculated by looking at the ratio of decay rates $\Gamma$ with different helicity,
\begin{equation}
\langle \Lambda \rangle = \frac{\Gamma(\mathbf{p}\cdot \boldsymbol{\xi} = |p|) - \Gamma(\mathbf{p}\cdot \boldsymbol{\xi}  = -|p|)}{\Gamma(\mathbf{p}\cdot \boldsymbol{\xi} = |p|) + \Gamma(\mathbf{p}\cdot \boldsymbol{\xi} = -|p|)}\,.
\end{equation}

The small scale of the interactions and the immense magnitude of the magnetic field make it necessary to consider the Landau levels the electrons decay into.  Detailed analysis of the effect of the Landau levels on interactions while considering the complete  electron wave function has concluded that the only significant change is in the energy of the electron, 
\begin{equation}
E_e^2 = \mathbf{p}_e^2 + m_e^2(1+2nb)\,, n=\begin{cases}1,2,...&\textrm{ spin up} \\ 0,1,2,...&\textrm{ spin down} \end{cases}
\end{equation}
where $b=B/B_{\textrm{c}}$ is the ratio of the magnetic field and the critical magnetic field, $B_{\textrm{c}} = \frac{m_e^2c^3}{\bar{h}e} \sim 4.4\cdot 10^{13}$ G.  

The magnetic field also affects the phase space of the electron.  It can only carry linear momentum in the direction of the magnetic field.  The magnetic field also causes degeneracy of levels.  With this taken in to account the phase space becomes,
\begin{equation}
\int \frac{dp_e}{2\pi} \frac{m_e^2 b}{2\pi} \,. 
\end{equation}

We will confine ourselves to calculating the helicity for an electron created through beta decay, one half of the direct Urca processes, Figure \ref{fig:feynman_direct}.   We sum over everything but the electron spin, 
\begin{eqnarray}
\sum |M|^2 \simeq 16 G_F^2 E_{\scriptscriptstyle{N}} E_{\scriptscriptstyle{P}}&&\left[ (1+3C_A^2) E_v(E_e -\mathbf{p}_e\cdot \boldsymbol{\xi}_e) \right. \\ \nonumber && \left. -(1-C_A^2) \mathbf{p_\nu\cdot (p}_e - m_e\mathbf{s}_e)\right]\,,
\end{eqnarray}
where $C_A\simeq1.26$ is the axial current constant, $G_F$ is the Fermi coupling constant, 
\begin{equation}
\mathbf{s}=\boldsymbol{\xi} + \frac{(\mathbf{p}\cdot\boldsymbol{\xi})\mathbf{p}}{m(m+E)}
\end{equation}
and $\boldsymbol{\xi}$ is the unit polarization vector. 

In order to find the decay rate we use the modified electron phase space.  To find the total decay rate it is necessary to sum over the probability of an electron appearing in the each of the Landau levels.  The sum truncates where the energy of a Landau level is high enough to be disallowed by conservation of energy. Specifically we let $n_{\textrm{max}}$ be the largest $n$ that satisfies $E_e^2 > m_e^2(1+2nb)$.  The total decay rate is the sum of the decay rate of each level,
\begin{equation}
\Gamma = \sum_{n=0}^{n_{\textrm{max}}} \Gamma_n\,,
\end{equation}  
where 
\begin{equation}
n_{\textrm{max}}= (E_e^2-m_e^2)/m_e^22b\,.  
\end{equation}  
The highest Landau level is reached when all of the energy goes into putting the electron in the highest landau level and none into the momentum $p_z$.  We see that increasing the magnetic field decreases the number of levels to which the electron has access.  When $b > E_e^2/2m_e^2 - 1/2$ the electron only has access to the lowest Landau level meaning that all the electrons are spin down.    

We must also account for the non-zero chemical potentials of the protons, electrons, and neutrons.   Because the electron and proton have large chemical potentials their phase spaces are constricted.  The presence of a  Fermi surface means that only higher energy electrons and protons can be created.  This is modelled by multiplying the usual phase space by Fermi blocking terms $1-f(E)$, where $f(E)$ is the Fermi distribution.   With these considerations the decay rate into a single landau level $n$ is, 
\begin{equation}
d\Gamma_n = \frac{1}{2E_{\scriptscriptstyle{N}}} \frac{dp_z}{2E_e2\pi} \frac{m_e^2 b}{2\pi} \frac{d^3p_\nu}{2E_\nu(2\pi)^3} \frac{d^3p_{\scriptscriptstyle{P}}}{2E_{\scriptscriptstyle{P}}(2\pi)^3}|M|^2(2\pi)^4\delta^4(p_{\scriptscriptstyle{N}}-p_{\scriptscriptstyle{P}}-p_e-p_\nu)(1-f_{\scriptscriptstyle{P}})(1-f_e)\,,
\end{equation}
There are a few quantities that naturally align themselves with the magnetic field.  Firstly, all spins are either aligned or anti-aligned so if we choose $\mathbf{B} = (0,0,B_z)$, then 
\begin{equation}
\boldsymbol{\xi}_e =(0,0,\xi_e)\textrm{ and } \boldsymbol{\xi}_{\scriptscriptstyle{N}} = (0,0,\xi_{\scriptscriptstyle{N}})\,.  
\end{equation}
Also, because of the Landau levels, the electrons have linear momentum only in the direction of the magnetic field, 
\begin{equation}
\mathbf{p}_e = (0,0,p_e)\,.
\end{equation}
The matrix element is reduced to
\begin{eqnarray}
\sum |M|^2 &\simeq& 8 G_F^2 E_{\scriptscriptstyle{N}} E_{\scriptscriptstyle{P}}\left[ (1+3C_A^2) E_v(E_e -p_e\xi_e) \right. \\ \nonumber && \left.-(1-C_A^2) |\mathbf{p}_\nu|(p_e - m_e s_e)\cos{\theta} \right]\,,
\end{eqnarray}
where $\theta$ is the angle between the z-axis and the direction of the neutrino momentum. 

The integrals up to the final electron integral are straight forward.  We are left with,
\begin{equation}
d\Gamma_n = A\, g_n\, dp_e  (E_{\scriptscriptstyle{N}} -E_{\scriptscriptstyle{P}}-E_e)^2\left[1-\frac{p_e\xi_e}{E_e} \right](1-f_e)\,,
\end{equation}
where,
\begin{equation}
A=\frac{m_e^2G_F^2b}{(2\pi)^3}(1-3C_A^2)(1-f_{\scriptscriptstyle{P}}(k_{\scriptscriptstyle{P}}))\,.
\end{equation}
The term $p_e\, \xi_e=\pm|p_e|$ gives us the helicity eigenstates.  We can get the combinations required for finding the average helicity, 
\begin{eqnarray}
\Gamma_n(+) - \Gamma_n(-) &=& -4Ag_n\int_0^{p_0} dp_e (E_{\scriptscriptstyle{N}} -E_{\scriptscriptstyle{P}}-E_e)^2\frac{|p_e|}{E_e} (1-f_e) \\  \Gamma_n(+) + \Gamma_n(-) &=& 4Ag_n\int_0^{p_0} dp_e (E_{\scriptscriptstyle{N}} -E_{\scriptscriptstyle{P}}-E_e)^2 (1-f_e)\,.
\end{eqnarray}
In order to do the final integral over the Fermi distribution we appeal to the Sommerfeld expansion,  
\begin{equation}
\int_0^{E_0} h(E)(1-f(E))dE = \int_\mu^{E_0}h(E)dE - \left. \frac{T^2\pi^2}{6} \frac{\partial}{\partial E}h(E)\right|_{E=\mu}\,.
\end{equation}
The particles in a neutron star are in chemical equilibrium.  The maximum energy available for the electron is equal to its chemical potential, $E_0 =\mu$, leaving the integral part of the Sommerfeld expansion to vanish.  Physically this makes sense because the equilibrium processes only occur thermally---the transition rate at zero temperature vanishes.  Doing our integral has been reduced to taking a derivative,  
\begin{eqnarray}
\Gamma_n(+) - \Gamma_n(-) &=& \left. \frac{2Ag_nT^2\pi^2}{3}\frac{p_e}{E_e}\frac{\partial}{\partial p_e}(E_{\scriptscriptstyle{N}} -E_{\scriptscriptstyle{P}}-E_e)^2\frac{p_e}{E_e}\right|_{p_e=k_e} \\
\Gamma_n(+) + \Gamma_n(-) &=& -\left. \frac{2Ag_nT^2\pi^2}{3}\frac{p_e}{E_e}\frac{\partial}{\partial p_e}(E_{\scriptscriptstyle{N}} -E_{\scriptscriptstyle{P}}-E_e)^2 \right|_{p_e=k_e}\,,
\end{eqnarray}
where remember that $E_e(p_e,n)$.  We can sum over each of these to get the total decay rate, then take the ratio to get the average helicity.  The important values are $E_{\scriptscriptstyle{N}} - E_{\scriptscriptstyle{P}} = \mu_e\sim k_e$ and the sum goes up to $n_{\textrm{max}} = k_e^2/2m_e^2b$.  The details after this are largely uninteresting and is computed by doing the sum numerically.  As a check we find that the helicity at zero magnetic field is $\langle \Lambda \rangle = -1$ as we expect.  Over the range of fields we are interested $B=10^{12}-10^{15}$ G the helicity is surprisingly constant.  We arrive at,
\begin{equation}
\label{helicity}
 \langle \Lambda \rangle = -0.84\,.
\end{equation}
The helicity is close to $-1$, but not so close that it doesn't warrant a comment.  With a large magnetic field the electron is forced into Landau levels. The single spin down state in the lowest level sometimes forces the electron into a right-handed configuration to conserve momentum. This occurs when the proton is created with the same spin as the initial neutron. In this case the electron and the neutrino must have then have opposite spins. If the electron is forced by the lowest Landau level to be spin down then the antineutrino must be spin up. Being right-handed, the antineutrino moves up in the direction of its spin. In order to conserve momentum the electron must move down, forming a right-handed configuration. The electron is bullied into being right-handed by the neutrino. The effect is not absolute though, and the intrinsic left-handedness of the weak interaction wins out.

\bibliographystyle{JHEP}
%\bibliography{mybib}

\providecommand{\href}[2]{#2}\begingroup\raggedright\endgroup

\end{document}